\documentclass[pra,twocolumn,showpacs,floatfix,superscriptaddress,aps]{revtex4}
\usepackage{amsmath}
\usepackage{makeidx}
\usepackage{amssymb}
\usepackage{graphicx}

\usepackage[usenames]{color}
\usepackage[normalem]{ulem}

\newcommand{\beq}{\begin{equation}}
\newcommand{\eeq}{\end{equation}} 
\newcommand{\beqa}{\begin{eqnarray}}
\newcommand{\eeqa}{\end{eqnarray}} 
\begin{document}

\title{Mixing, demixing, and structure formation in a binary dipolar Bose-Einstein condensate}

\author{Luis E. Young-S.\footnote{lyoung@ift.unesp.br} }
\author{ S. K. Adhikari\footnote{adhikari@ift.unesp.br; URL:
http://www.ift.unesp.br/users/adhikari}}
\affiliation{
Instituto de F\'{\i}sica Te\'orica, UNESP - Universidade Estadual Paulista, \\ 01.140-070 S\~ao Paulo, S\~ao Paulo, Brazil
} 

\begin{abstract}

We study static properties of disk-shaped  binary dipolar Bose-Einstein condensates   of  $^{168}$Er-$^{164}$Dy and $^{52}$Cr-$^{164}$Dy mixtures
under the action of inter- and intra-species contact and dipolar interactions
and demonstrate the effect of dipolar interaction using the mean-field approach.
 Throughout this study we use realistic values of 
inter- and intra-species dipolar interactions and the intra-species scattering lengths and consider the   inter-species scattering length as a parameter. The stability of the binary mixture is illustrated through phase plots involving number of atoms of the species. The binary   system always becomes unstable as the number of atoms increases beyond a certain limit. As the inter-species scattering length increases corresponding to more repulsion, an overlapping  mixed state of  
the two species changes to a separated demixed configuration.  During transition from a mixed to a demixed configuration as the inter-species scattering length is increased for parameters just below the stability line, the binary condensate 
shows special structures in density in the form of red-blood-cell-like biconcave and Saturn-ring-like shapes, which are direct manifestations of dipolar interaction.

\end{abstract}

\pacs{03.75.Hh, 03.75.Mn}

\maketitle

\section{Introduction}

After the experimental realization \cite{ExpCr,cr,saddle,crrev,52Cr} of a dipolar Bose-Einstein condensate 
(BEC) of $^{52}$Cr atoms with magnetic moments, there has been renewed interest in the study 
of static and dynamic properties of such a condensate in the pursuit of 
novel and interesting properties and features emerging as a consequence 
of anisotropic long-range dipolar interaction. The atomic interaction in a dilute 
BEC of alkali-metal atoms is taken as a S-wave short-range 
(delta-function) potential. However, the anisotropic long-range dipolar interaction 
is nonlocal in nature acting in all partial waves. More 
recently, BEC of $^{164}$Dy \cite{ExpDy,dy} and $^{168}$Er \cite{ExpEr} atoms with larger dipole 
moments are available for experimental studies, and polar molecules 
with much larger (electric) dipole moment are being considered \cite{polar} for BEC 
experiments. Among the novel features of a BEC with anisotropic 
dipolar interaction, one can mention the peculiar shape  and stability properties
of a stationary 
state \cite{shape}, red-blood-cell-like biconcave shape in density due to radial and angular roton-like excitations 
\cite{roton}, anisotropic 
D-wave collapse \cite{collapse}, formation of anisotropic soliton, vortex soliton \cite{soliton} and 
vortex lattice \cite{lattice}, 
anisotropic sound 
and shock wave propagation \cite{shock}, and anisotropic Landau critical velocity \cite{landau}
among others. Distinct stable
checkerboard, stripe, and star configurations in dipolar
BECs have been identified in a two-dimensional (2D) optical
lattice as stable Mott insulator \cite{mott} as well as superfluid
soliton \cite{15} states.
A new possibility
of studying universal properties of dipolar BECs
for large scattering length has been suggested \cite{unitarity}.
Many of these features have only been predicted 
theoretically, although some of them have already been experimentally 
confirmed. To enhance the effect of the anisotropic dipolar interaction 
in most of these theoretical studies, the contact interaction has been 
set equal to zero. In $^{52}$Cr atom, this could be necessary as the 
strength of the repulsive contact interaction is much stronger than that 
of the dipolar interaction. However, dipolar BEC of atoms such as 
$^{164}$Dy with much larger dipolar interaction can show effects of 
anisotropic nonlocal dipolar interaction  without switching off the contact interaction.

Now with available dipolar BECs of different  species of atoms, there is the 
possibility to investigate a binary mixture with two dipolar BECs of two types
of dipolar interactions acting on each species, e.g., intra- and 
inter-species, superposed on intra- and inter-species contact interactions, 
thus creating a much richer platform to study the effect of dipolar 
interaction. { Previously, there have been studies of nondipolar 
binary boson-boson \cite{bb,bb2}, boson-fermion \cite{bf} and fermion-fermion \cite{ff} 
as well as dipolar-nondipolar binary boson-boson \cite{dnd}
mixtures.} 
For a binary system
without dipolar interactions it is possible to  have either a mixed
or demixed phase.
 In this paper, we consider binary dipolar BECs composed of
$^{164}$Dy and $^{52}$Cr atoms as well as $^{164}$Dy and $^{168}$Er  atoms. 
With all the dipole moments aligned along the $z$ axis, the net dipolar interaction is attractive in the cigar shape along $z$ axis
and repulsive in the disk shape confined in the $x-y$ plane,
as parallel dipoles in a chain along $z$ axis attract and those confined in the $x-y$ plane repeal each other. Consequently, the binary dipolar BEC 
with dipoles polarized along $z$ axis is more stable in the disk shape confined in the $x-y$ plane compared to the cigar shape along the $z$ axis
vulnerable to collapse
and we shall consider only such a  disk-shaped binary dipolar BEC in this study. Even such a  disk-shaped binary dipolar BEC is found to be
unstable due to collapse instability for dipolar interaction above a critical value controlled by the number of atoms.  This is consistent with a similar conclusion that a single-component dipolar BEC for any trap anisotropy collapses above a critical value 
of dipolar interaction \cite{bohn,17}.

We study the limits of stability of binary dipolar disk-shaped BECs 
formed of $^{168}$Er and $^{164}$Dy atoms and of $^{52}$Cr and 
$^{164}$Dy atoms and present the results in terms of stability phase 
plots which should aid in the experimental preparation and study of 
binary dipolar BECs.  Of all the atomic interactions, the inter- and 
intra-species dipolar interactions controlled by the known dipole 
moments of the two species will be considered known. The contact 
interactions governed by the approximately known intra-species 
scattering lengths will also be taken as fixed parameters in this study. 
We then study the stability of the binary mixture by varying the number 
of atoms in each species and the yet unknown inter-species scattering 
length $a_{12}$. For small values of $a_{12}$, the dipolar binary BEC 
prefers a mixed configuration and with the augmentation of the 
inter-species repulsion the system moves into a demixed configuration. 
During this process of mixing-demixing, distinct biconcave 
red-blood-cell-like and Saturn-ring-like shapes are found in the 
densities of the two components just below the line of stability of the 
binary dipolar BEC for a certain value of inter-species scattering 
length. This configuration corresponds to saddle-point structures in 
the 2D densities of the two components in the $x-z$ plane. It was 
demonstrated \cite{saddle} that the dipolar interaction corresponds to a 
saddle structure in 2D responsible for the saddle structure in density 
in the present study on binary dipolar BEC. The biconcave structure is a 
consequence of roton instability near the stability line due to dipolar 
interaction and was studied in detail in single-component dipolar BECs 
\cite{roton}.  The Saturn-ring-like density distribution was never 
observed in a single-component dipolar BEC. To enhance the dipolar 
effect, in these previous studies the repulsive contact interaction was 
set to zero, whereas in the present study we set all the intra- and 
inter-species repulsive contact interactions to their large experimental 
values and yet obtain these special structures because of the larger 
values of dipole interactions appropriate for $^{164}$Dy and $^{168}$Er 
atoms.

In Sec. II we present the mean-field model for the binary dipolar BEC 
interacting via inter- and intra-species contact and dipolar 
interactions in an axially-symmetric confinement. Details of numerics 
together with an exposition of our numerical results are included in 
Sec. III. After the application of our model to a simplified binary 
dipolar BEC without any contact interaction, we present the results for 
the binary $^{168}$Er-$^{164}$Dy and $^{52}$Cr-$^{164}$Dy mixtures. 
First, we obtain the stability phase plots of the binary dipolar BECs. 
Then we demonstrate mixing, demixing, and structure formation in 
densities in both cases.  Close to the stability line, we find 
Saturn-ring- and red-blood-cell-like density distributions in the 
components $-$ only possible in the presence of dipolar interaction. 
Finally, in Sec. III we present a brief summary of our findings.

\section{Mean-field model for the binary dipolar BEC}

We consider a general two-species dipolar BEC with inter- and intra-species dipolar and contact interactions with  the mass, number, magnetic moment, scattering length denoted by $m_i, N_i, \mu_i, a_i, i=1,2$, respectively. 
The angular frequencies for the axially-symmetric trap along $x$, $y$ and $z$ directions are taken as $\omega_x^{(i)}=\omega_y^{(i)}=\omega_i$ and $\omega_z=\lambda_i\omega_i$. The intra- and inter-species interactions for two atoms at
positions $\bf r_1$ and $\bf r_2$ are  
taken as 
\begin{eqnarray}\label{intrapot}
V_i({\bf R})= \frac{\mu_0\mu_i^2}{4\pi}\frac{1-3\cos^2 \theta}{|{\bf R}|^3}+\frac{4\pi \hbar^2 a_i}{m_i}\delta({\bf R }),\\
\label{interpot}
V_{12}({\bf R})= \frac{\mu_0\mu_1\mu_2}{4\pi}\frac{1-3\cos^2 \theta}{|{\bf R}|^3}+\frac{2\pi \hbar^2 a_{12}}{m_R}\delta({\bf R}),
     \end{eqnarray}
where $\bf R = r_1-r_2,$ $\mu_0$ is the permeability of free space, $\theta$ is the angle made 
by the vector ${\bf R}$ with the polarization $z$ direction, 
  and $m_R=m_1m_2/(m_1+m_2)$ is the reduced mass of the two species of atoms. 
With these interactions,  
the coupled Gross-Pitaevskii (GP) equations for the binary dipolar BEC can be written as \cite{crrev}
\begin{align}& \,
i \hbar \frac{\partial \phi_1({\bf r},t)}{\partial t}=
{\Big [}  -\frac{\hbar^2}{2m_1}\nabla^2
+ \frac{1}{2}m_1 \omega_1^2 
(\rho^2+\lambda^2_1{z}^2 )\nonumber
\\ 
& 
+ N_1 \frac{ \mu_0 \ {\mu}^2_1 }{4\pi}
\int V_{dd}({\mathbf R})\vert\phi_1({\mathbf r'},t)\vert^2 d{\mathbf r}'  
 \nonumber \\  &
+\frac{2\pi \hbar^2}{m_R} {a}_{12} N_2 \vert \phi_2({\bf r},t)|^2 
+ \frac{4\pi \hbar^2}{m_1}{a}_1 N_1 \vert \phi_1({\bf r},t)\vert^2
\nonumber 
\\ &
+ N_2 \frac{ \mu_0 \ {\mu}_1 \ {\mu}_2}{4\pi}
\int V_{dd}({\mathbf R})\vert\phi_2({\mathbf r'},t)\vert^2 d{\mathbf r}' 
{\Big ]}  \phi_1({\bf r},t),
\label{eq1}
\\
\label{eq2}
&i \hbar \frac{\partial \phi_2({\bf r},t)}{\partial t}=
{\Big [}  -\frac{\hbar^2}{2m_2}\nabla^2
+ \frac{1}{2}m_2 \omega_2^2 \ 
(\rho^2+\lambda^2_2{z}^2 )
\nonumber\\ &  
+ N_2 \frac{ \mu_0 \ {\mu}^2_2 }{4\pi}
\int V_{dd}({\mathbf R})\vert\phi_2({\mathbf r'},t)\vert^2 d{\mathbf r}' \nonumber 
\\ & 
+\frac{2\pi \hbar^2}{m_R} {a}_{12} N_1 \vert \phi_1({\bf r},t) \vert^2
+ \frac{4\pi \hbar^2}{m_2}{a}_2 N_2 \vert \phi_2({\bf r},t) \vert^2 
\nonumber \\
&
+ N_1 \frac{ \mu_0 \ {\mu}_1 \ {\mu}_2}{4\pi}
\int V_{dd}({\mathbf R})\vert\phi_1({\mathbf r'},t)\vert^2 d{\mathbf r}' 
{\Big ]}  \phi_2({\bf r},t),
\\&
V_{dd}({\mathbf R})= 
\frac{1-3\cos^2\theta}{{\mathbf R}^3},
\end{align}
with  $\rho^2=x^2+y^2$.  

To compare the dipolar and contact interactions, the intra- and inter-species dipolar interactions will be  expressed in terms the length scales
$a^{(i)}_{dd}$ and $a^{(12)}_{dd}$, respectively, defined by
\begin{align}
\frac{\mu_0\mu_i^2}{4\pi}= \frac{3\hbar^2}{m_i} a^{(i)}_{dd}, \quad
 \frac{\mu_0\mu_1\mu_2}{4\pi}= \frac{3\hbar^2}{m_1} a^{(12)}_{dd}.
\end{align}
For the intra-species dipolar scale $a^{(i)}_{dd}$, the mass of the corresponding species $m_i$ has been used to define the scale. To define the inter-species dipolar scale $a^{(12)}_{dd}$, we have used the mass $m_1$ of the first species. We express the strengths of the dipolar interactions in Eqs. (\ref{eq1}) and (\ref{eq2}) by these length scales 
and transform these equations into the following dimensionless form:
\begin{align}& \,
i \frac{\partial \phi_1({\bf r},t)}{\partial t}=
{\Big [}  -\frac{\nabla^2}{2 }
+ \frac{1 }{2} (\rho^2+\lambda^2_1 z^2 ) \nonumber  \\ &
+ g_1 \vert \phi_1 \vert^2 
+ g_{dd}^{(1)}
\int V_{dd}({\mathbf R})\vert\phi_1({\mathbf r'},t)\vert^2 d{\mathbf r}'  \nonumber\\ & \, %
+ g_{12} \vert \phi_2 \vert^2
+ g_{dd}^{(12)}
\int V_{dd}({\mathbf R})\vert\phi_2({\mathbf r'},t)\vert^2 d{\mathbf r}' 
{\Big ]}  \phi_1({\bf r},t),
\label{eq3}\\
& \,
i \frac{\partial \phi_2({\bf r},t)}{\partial t}={\Big [}  
-m_{12} \frac{\nabla^2}{2}
+m_w \frac{1}{2}(\rho^2+\lambda^2_2 z^2 )\nonumber  \\ &
+ g_2 \vert \phi_2 \vert^2 
+ g_{dd}^{(2)}
\int V_{dd}({\mathbf R})\vert\phi_2({\mathbf r'},t)\vert^2 d{\mathbf r}'  \nonumber \\ & \,
+ g_{21} \vert \phi_1 \vert^2
+ g_{dd}^{(21)}
\int V_{dd}({\mathbf R})\vert\phi_1({\mathbf r'},t)\vert^2 d{\mathbf r}' 
{\Big ]}  \phi_2({\bf r},t),
\label{eq4}
\end{align}
where
$m_{12}={m_1}/{m_2},$
$m_w={ \omega_2^2}/{(m_{12}\omega_1^2)},$
$g_1=4\pi a_1 N_1,$
$g_{dd}^{(1)}=3N_1 a_{dd}^{(1)},$
$g_2= 4\pi a_2 N_2 m_{12},$
$g_{12}={2\pi m_1} a_{12} N_2/m_R,$
$g_{21}={2\pi m_1} a_{12} N_1/m_R,$
$g_{dd}^{(2)}= 3N_2 a_{dd}^{(2)}m_{12},$
$g_{dd}^{(12)}=3N_2  a_{dd}^{(12)},$
$g_{dd}^{(21)}=3N_1 a_{dd}^{(12)}. $
In Eqs. (\ref{eq3}) and (\ref{eq4}), length is expressed in units of oscillator length for the first species $l_0=\sqrt{\hbar/m_1\omega_1}$, energy in units of oscillator energy $\hbar\omega_1$, density $|\phi_i|^2$ in units of $l_0^{-3}$, and 
time in units of $ t_0=\omega_1^{-1}$.

\section{Numerical Results}

For the binary dipolar BEC  we solve Eqs. (\ref{eq3}) and (\ref{eq4}) numerically after discretization \cite{CPC}. The divergence of the dipolar term at short distances 
has been handled by treating this term in momentum ($\bf k$) space. The dipolar integral 
in (Fourier) momentum space is tackled by the following convolution integral 
\cite{Santos01,crrev}
\begin{align}&
\int d{\bf r'}V_{dd}({\bf R})n({\bf r'})
=\int \frac{d{\bf k}}{(2\pi)^ 3}e^{-i{\bf k \cdot r}}
 V_{dd}({\bf k})n({\bf k}),  
\end{align}
with $n({\bf r})=|\phi({\bf r})|^2$.
The Fourier transformation (FT) is defined by 
\begin{align}
&
A({\bf k})=\int d{\bf r} B({\bf r})e^{i {\bf k \cdot r}},
\\&
 B({\bf r})=\frac{1}{(2\pi )^3}
\int d{\bf k} A({\bf k})e^{-i {\bf k \cdot r}}.
\end{align}
 The FT $V_{dd}({\bf k})$ of the 
dipolar potential is known analytically \cite{crrev,Santos01}:
\begin{align}
V_{dd}({\bf k})=\frac{4\pi}{3} \left(\frac{3k_z^2}{{\bf k}^2} -1   \right).
\end{align}
 The FT $n({\bf k})$ of density 
is calculated numerically by a fast FT (FFT) routine.  The inverse FT is 
also evaluated numerically by the FFT routine. The whole procedure is 
performed in three-dimensional
(3D) Cartesian coordinate system irrespective of the underlying 
trap symmetry.

\subsection{Same species of atoms 
in isotropic trap}

\begin{figure}[!t]

\begin{center}
\includegraphics[width=\linewidth]{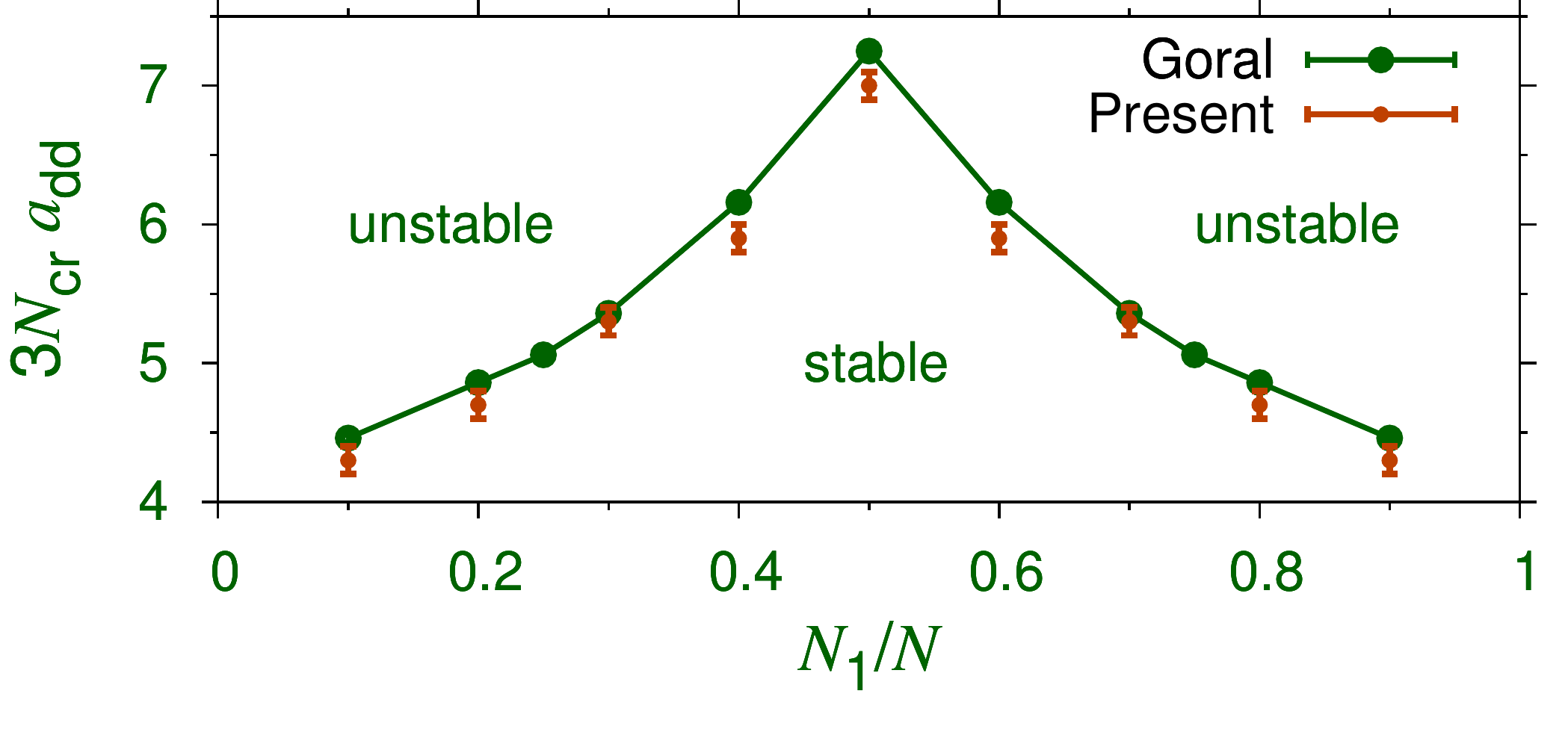}
\caption{ (Color online) The maximum allowed value 
of 
net intra-species
dipolar nonlinearity   $\zeta_{\mathrm{cr}} (\equiv 3N_{\mathrm{cr}}a_{dd})$ versus fraction of atoms 
of the first component $\eta$ $(\equiv N_1/N)$
  for a binary dipolar BEC  of oppositely
polarized dipolar gases in a spherically symmetric trap compared with the 
findings of G\'oral and Santos \cite{Santos01}.  }\label{fig1}
\end{center}

\end{figure}

To test the model and the numerical routine, we consider a simple binary dipolar BEC of same species of atoms without contact interaction
in a spherically-symmetric trap. However, the two components are considered to be oppositely polarized along $z$ and $-z$ directions, 
respectively. In this model, first considered by G\'oral and Santos 
 \cite{Santos01},
$m_i=m, \ \omega_i=\omega , 
a_{dd}^{(i)}=-a_{dd}^{(12)}=a_{dd},  
a_i=a_{12}=0 , \ m_{12}= m_\omega=1,  \lambda_{i}=1$.
%
Consequently, in  Eqs. (\ref{eq3}) and (\ref{eq4}) $g_i=g_{12}=0,
g_{dd}^{(i)}=3N_ia_{dd},  g_{dd}^{(12)}=-3N_2a_{dd},$ and $g_{dd}^{(21)}=-3N_1a_{dd}$. 
Using
$
\eta=N_1/N,  \zeta= 3a_{dd}N,  N=N_1+N_2$
we can write Eqs. (\ref{eq3}) and (\ref{eq4}) as in Ref. \cite{Santos01}:
\begin{align}& 
i \frac{\partial \phi_1}{\partial t}=
{\Big [}  -\frac{\nabla^2}{2 }
+ \frac{r^2 }{2}
+  \zeta {\Big \{} \eta
U_1({\mathbf r})
-(1-\eta)
U_2 ({\mathbf r}) {\Big \}}
{\Big ]}  \phi_1,
\label{eq5}\\
&
i \frac{\partial \phi_2}{\partial t}={\Big [}  
- \frac{\nabla^2}{2}
+ \frac{r^2 }{2}
+ \zeta {\Big \{} (1-\eta) U_2 ({\mathbf r})
- \eta   U_1 ({\mathbf r})
 {\Big \}}
{\Big ]}  \phi_2,
\label{eq6}\\
& U_i({\bf r})=\int V_{dd}({\mathbf R})\vert\phi_i({\mathbf r'})\vert^2 d{\mathbf r}', \quad i=1,2.
\end{align} 
For a fixed fraction of the atoms of the first component $\eta$, the system is stable up to a critical value of net dipolar interaction nonlinearity $\zeta$,
e.g. for $\zeta<\zeta_{\mathrm{cr}}$,
 beyond which the system becomes unstable. 
The stability of the binary system can be expressed in terms of a $\zeta-\eta$ phase diagram. 
This stability phase diagram for this  binary dipolar condensate of  same species of atoms, but  of oppositely polarized 
dipolar moments,  in a spherically symmetric trap is shown in Fig. \ref{fig1} in  agreement with G\'oral and Santos   \cite{Santos01}.
Independent of the composition of the system governed by the parameter $\eta$,
the system is unstable beyond a critical value of the total number of atoms $N$.  

\subsection{Different species of atoms in disk trap}

{{\bf Binary} $^{168}$Er-$^{164}$Dy {\bf mixture:}} We consider two 
different binary mixtures of dipolar BECs. First, we consider the 
$^{168}$Er-$^{164}$Dy mixture. In this case we take BEC number 1 as 
$^{168}$Er and BEC number 2 as $^{164}$Dy with parameters $\mu_1 
=7\mu_B, \mu_2=10\mu_B,  \ a_{dd}^{(1)}=66a_0, 
a_{dd}^{(12)}=94 a_0, a_{dd}^{(2)}=131a_0$ with $a_0$ the Bohr radius 
and $\mu_B$ the Bohr magneton. Without accurate experimental estimates 
\cite{ExpDy,ExpEr} of the intra-species scattering lengths, we use $a_i= 
110 a_0$. The angular frequencies of the axial trap for $^{164}$Dy are 
taken as $\omega_2= 2\pi \times{  243}$ Hz, $\lambda_2=\sqrt{10}\approx 
3.1623,$ compared with experimental frequencies \cite{ExpDy} 
$\{f_x,f_y,f_z \} =\{205,195,760\}$ Hz with trap aspect ratio $\lambda 
=3.8$. We take $m_\omega= (m_2 \omega_2^2)/(m_1\omega_1^ 2)=1$, 
corresponding to $\omega_1=2\pi \times {240}$ Hz, $\lambda_1= 
\sqrt{10}.$ The unit of length in this study is 
$l_0=\sqrt{\hbar/(m_1\omega_1)}\approx {0.5}$ $\mu$m and the unit of time 
$t_0=\omega_1^{-1}\approx {0.663}$ ms.


{{\bf Binary} $^{52}$Cr-$^{164}$Dy {\bf mixture:}} In this case, taking 
the condensate number 1 as $^{52}$Cr and the one number 2 as $^{164}$Dy, 
the parameters are: $\mu _1= 6\mu_B, \mu_2=10\mu_B,$ 
$a_{dd}^{(1)}=15a_0, 
a_{dd}^{(2)}=131 a_0, a_{dd}^{(12)}=25 a_0. $ We take $a_i=110 a_0,$ 
close to their experimental estimates \cite{ExpDy,ExpCr,52Cr}. The 
angular frequencies for the axial trap for $^{52}$Cr are taken as 
$\omega_1=2\pi \times {195}$ Hz, $\lambda_1=\sqrt{10}\approx 
3.1623$, compared with the experimental frequencies \cite{ExpCr} 
$\{f_x,f_y,f_z\}=\{150,150,\lambda f_x\}$ Hz with trap aspect ratio 
$\lambda< 10.$ We take $m_\omega= (m_2 \omega_2^2)/(m_1\omega_1^ 2)=1$, 
corresponding to $\omega_2=2\pi \times {110}$ Hz, $ \lambda_2=\sqrt{10}$. 
The unit of length in this calculation is 
$l_0=\sqrt{\hbar/(m_1\omega_1)}\approx {1}$ $\mu$m and the unit of time 
$t_0=\omega_1^{-1}\approx {0.816}$ ms.

\begin{figure}[!t]

\begin{center}
\includegraphics[width=\linewidth]{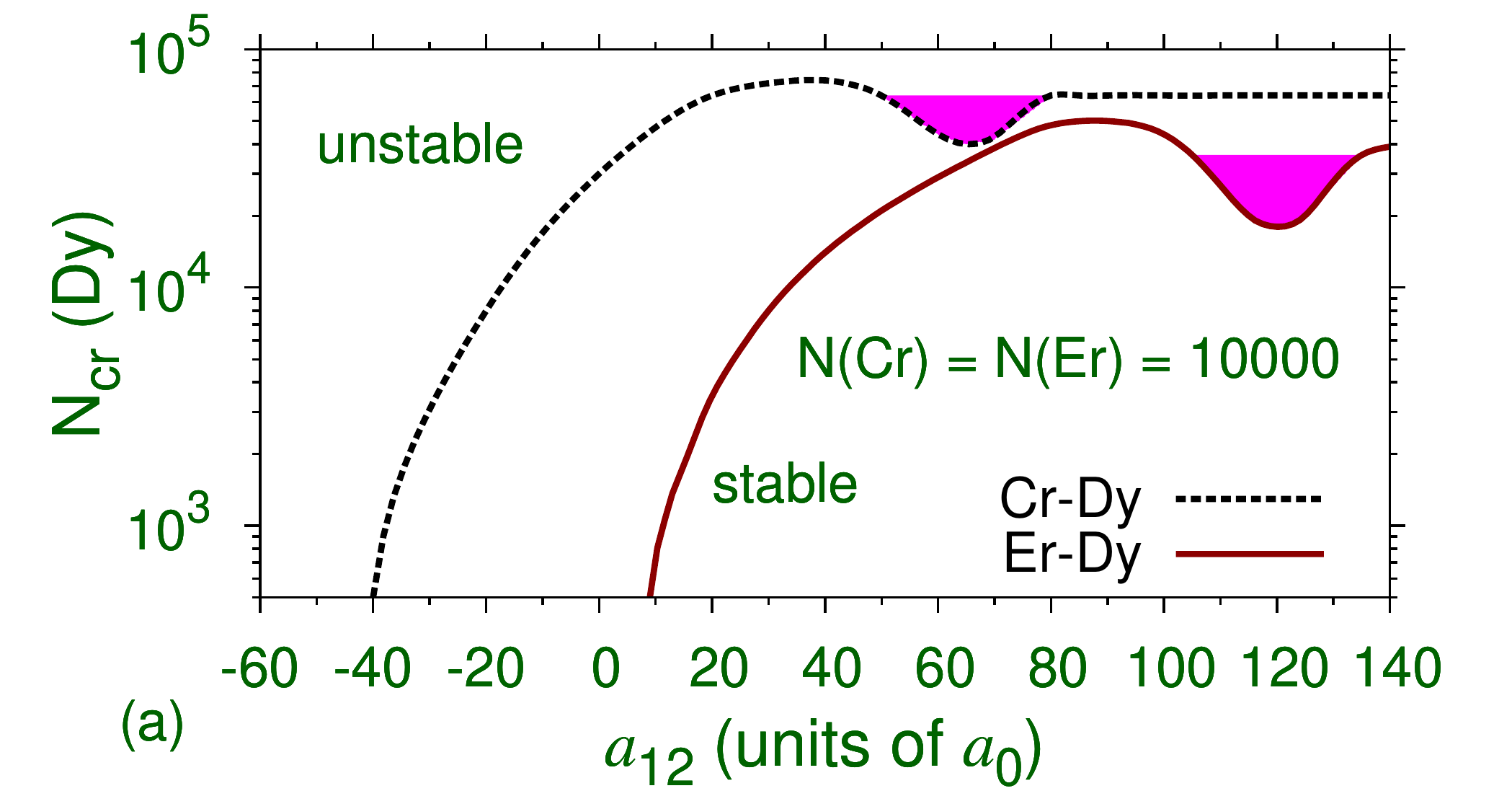}
\includegraphics[width=\linewidth]{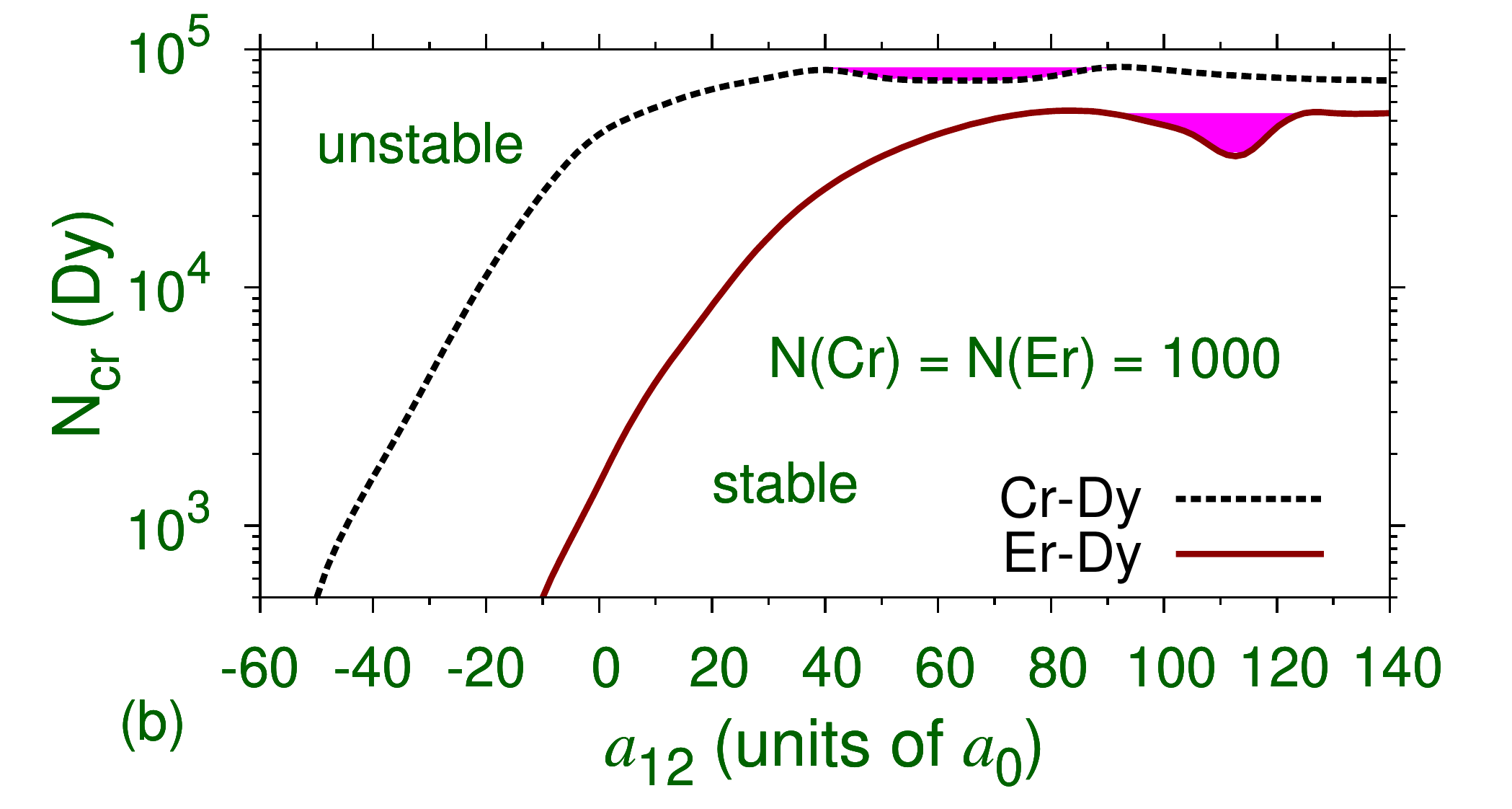}
\caption{(Color online) Stability phase plot  showing the critical number   
$N_{\mathrm cr}$(Dy) of  $^{164}$Dy atoms in the $^{52}$Cr-$^{164}$Dy
and $^{168}$Er-$^{164}$Dy binary mixtures versus inter-species scattering 
length $a_{12}$
for (a)  10000 and (b) 1000
$^{52}$Cr or
$^{168}$Er atoms, respectively. 
The system is stable below the respective 
lines. The shaded dark areas illustrate the  domains, where
biconcave 
density profiles appear.  
}\label{fig2}
\end{center}
\end{figure}

\subsection{Stability phase plot}

By solving the coupled set of GP equations (\ref{eq3}) and (\ref{eq4})
for the binary 
mixture of dipolar BECs  with the above parameters
we find that the  system becomes unstable above a certain 
number of   atoms with the total dipolar interaction beyond a limiting value in agreement with a similar conclusion \cite{bohn}
in a single-component  dipolar 
BEC.  
To perform a systematic study of the stability of the binary 
dipolar BEC,
we solve Eqs. 
(\ref{eq3}) and (\ref{eq4}) with the above parameters, fixing the number of atoms in the first species ($^{52}$Cr or $^{168}$Er) and searching for the critical number of atoms ($N_{\mathrm cr}$) of second species $^{164}$Dy
beyond which the system becomes unstable
for different values of inter-species scattering length $a_{12}$.
If the binary mixture has a smaller number of atoms of the first species 
($^{52}$Cr or $^{168}$Er), it can accommodate a larger number of atoms of the 
second species ($^{164}$Dy) while the limiting value of the total dipolar interaction is reached maintaining all other parameters fixed.

We present the results of our study in stability phase plots in Fig. \ref{fig2}.
In this figure  we show the critical number of $^{164}$Dy 
atoms $N_{\mathrm cr}$(Dy) versus $a_{12}$ for 10000 $^{52}$Cr or 
$^{168}$Er atoms in binary $^{52}$Cr-$^{164}$Dy and 
$^{168}$Er-$^{164}$Dy mixtures, respectively.  The system is stable below the 
lines of Fig. \ref{fig2} and unstable above.  
Stability phase plots of Fig. \ref{fig2}  could 
be relevant for planning future experiments on binary dipolar BEC. 
The system may have distinct structure in density  in the shaded regions in this figure.
This region is similar to the darker region in Fig.  \ref{fig1} of Ref. \cite{bohn}, where   biconcave shape in density of a single-component 
dipolar BEC
appeared due to roton instability. { The biconcave shape in density in a disk-shaped dipolar BEC
emerges because of  dipolar repulsion  in the plane of the disk. Due to this repulsion the atoms come to the peripheral region of the disk and 
a region of low density 
appears in the center. Similar low-density central region may appear in a rotating BEC 
due to centrifugal repulsion, as has been found in a binary nondipolar BEC in a rotating trap \cite{bb2}. } 
We shall see below that 
biconcave and Saturn-ring-like 
shapes in density  of the components of the binary dipolar BEC
may appear in the shaded region in  
Fig. \ref{fig2}. The system can accommodate a  larger number of atoms for large positive values of $a_{12}$ responsible for large contact repulsion which stabilize the binary dipolar BEC.  As $a_{12}$ reduces, the contact repulsion reduces 
and the system becomes more vulnerable to collapse for larger number of atoms due to dipolar interaction and hence can accommodate only a small number of 
$^{164}$Dy atoms as can be seen in Fig. \ref{fig2}.

\begin{figure}[!t]

\begin{center}

\includegraphics[width=\linewidth]{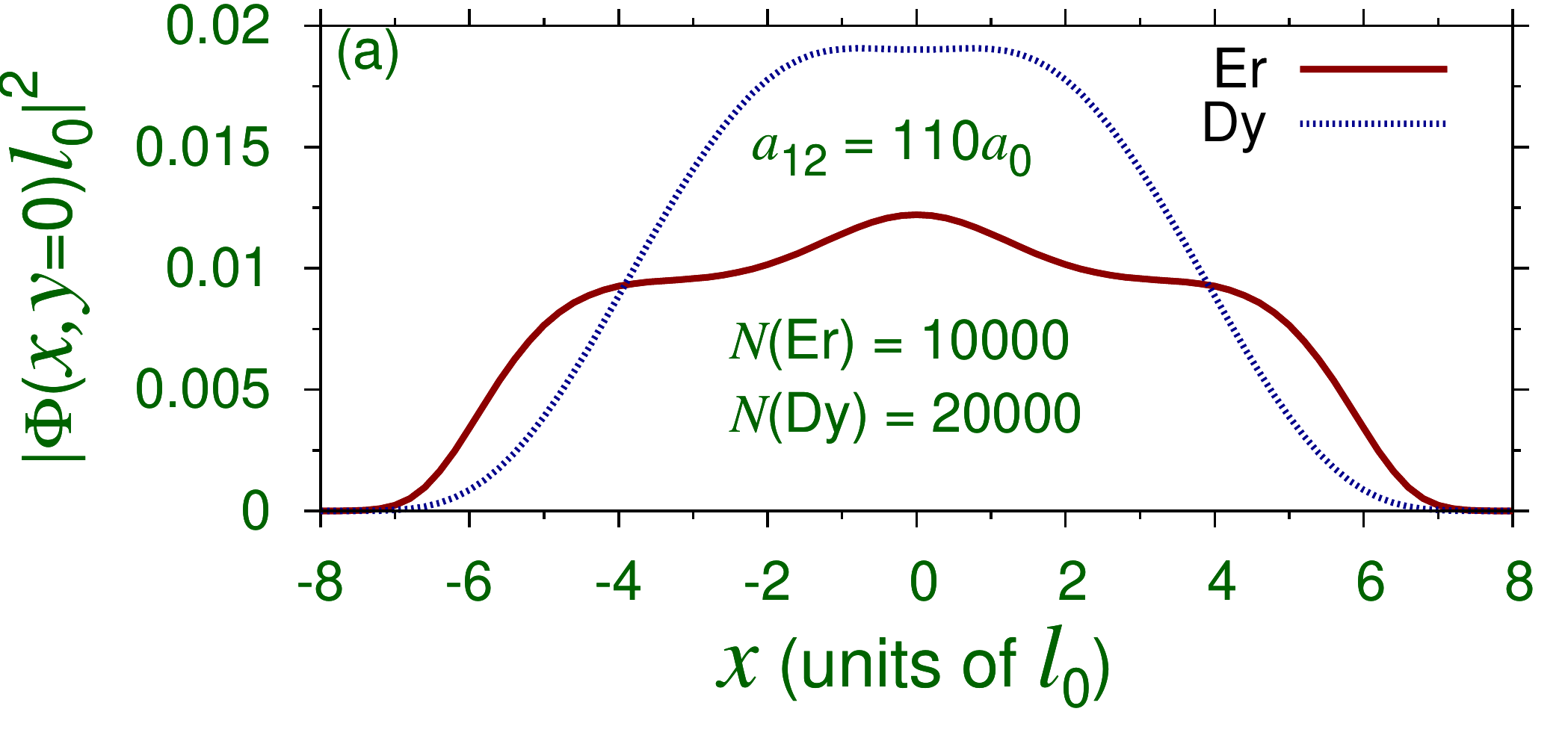}
\includegraphics[width=\linewidth]{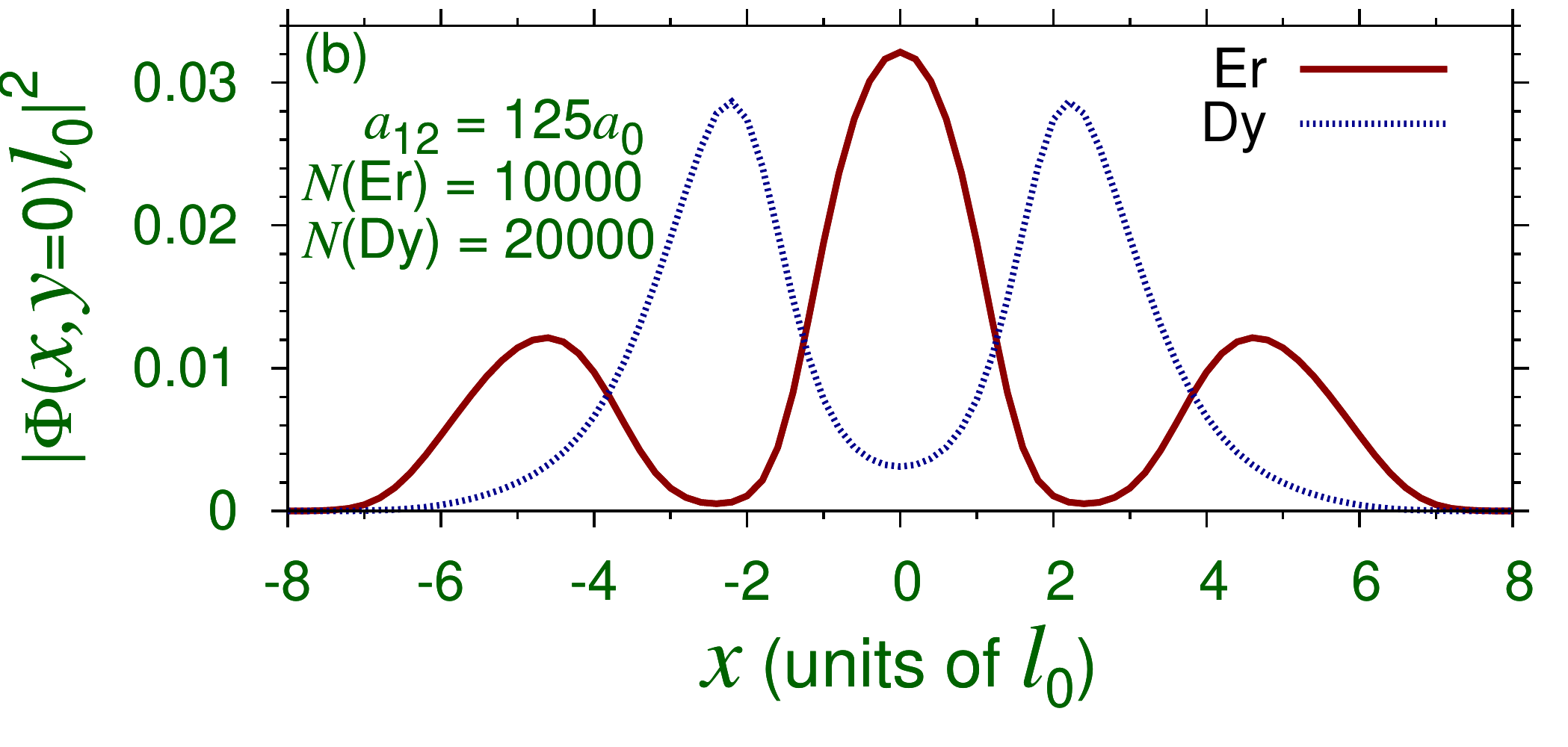}
\includegraphics[width=\linewidth]{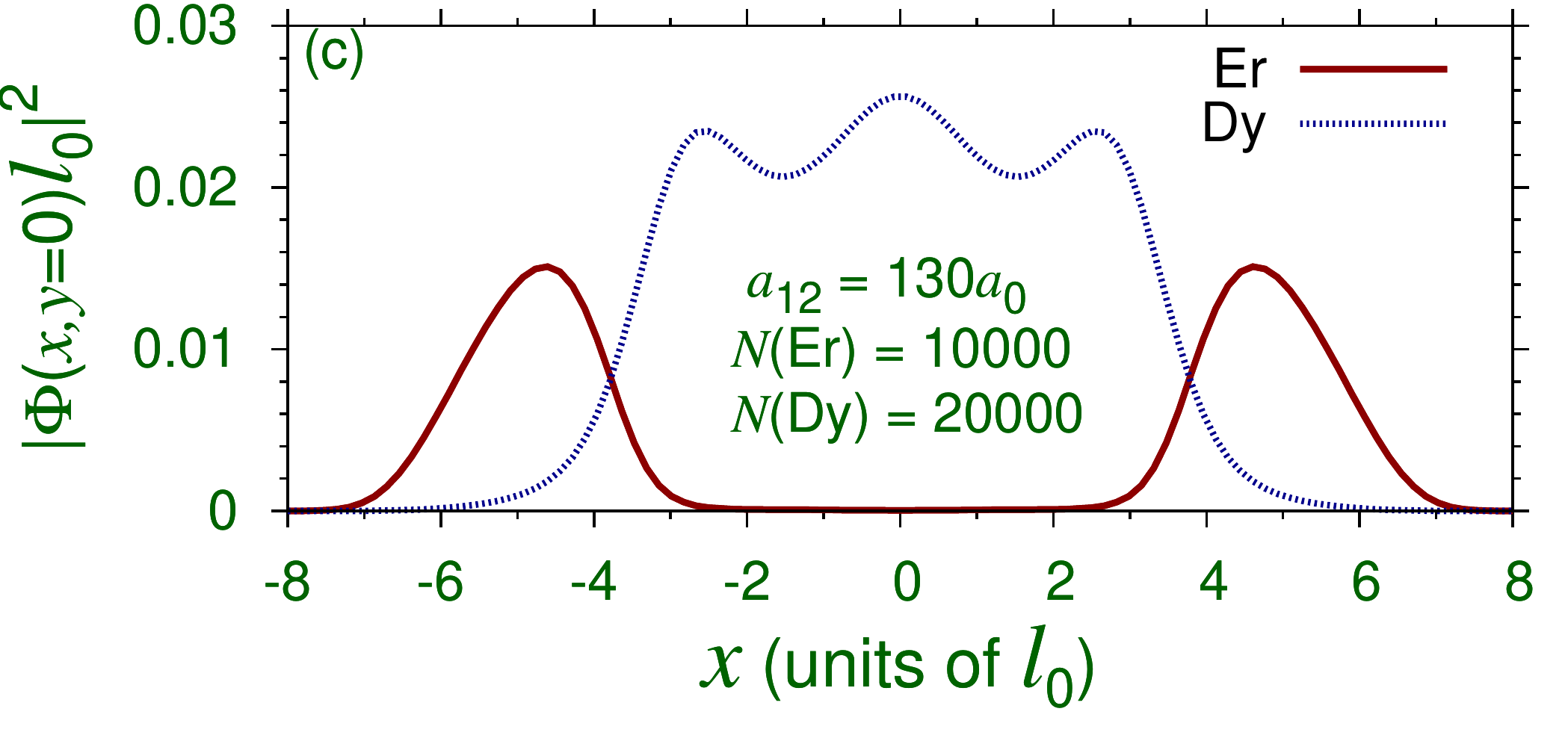}
\includegraphics[width=\linewidth]{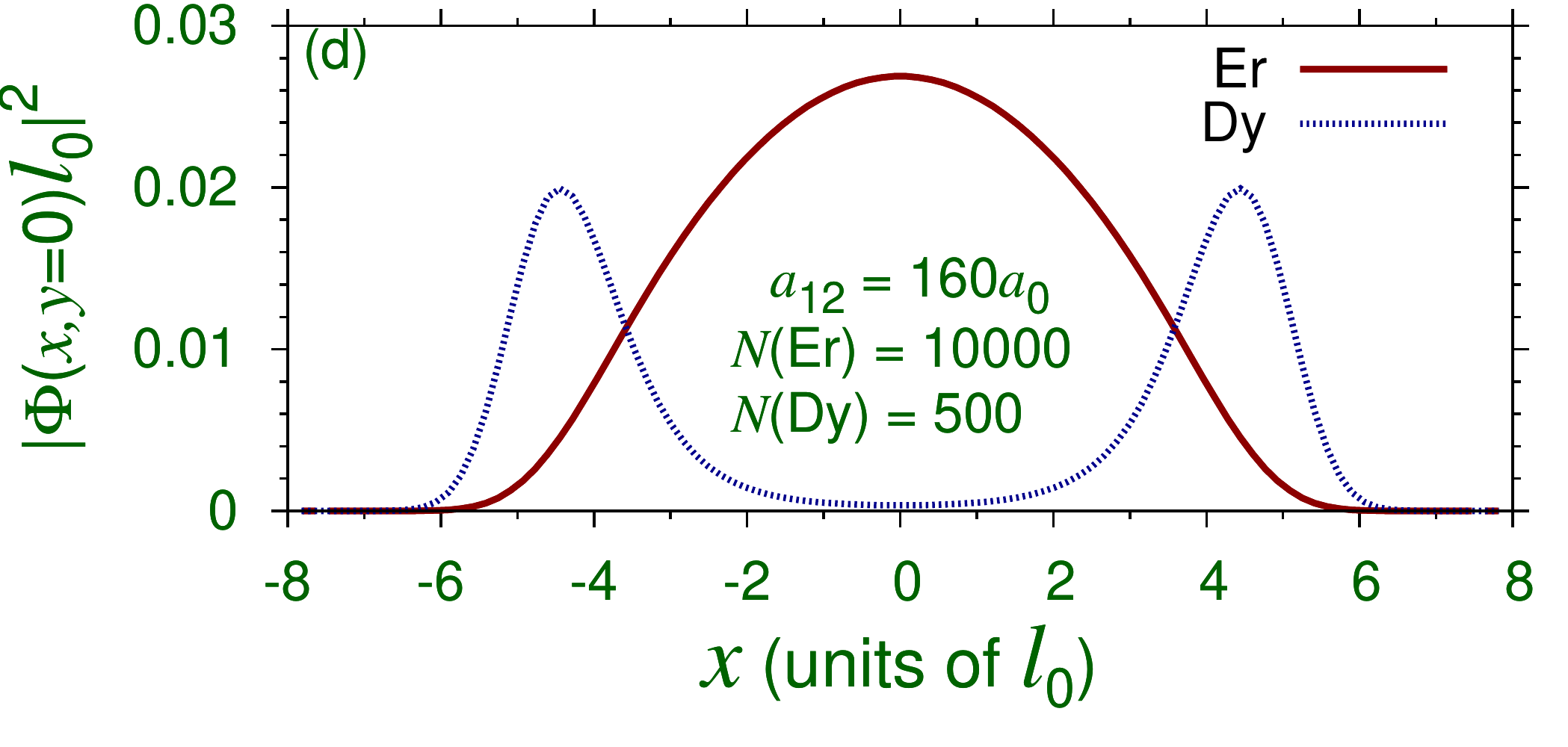}

\caption{(Color online) Two-dimensional 
 radial density along  $ x$ axis $|\Phi(x, y = 0)|^2 \equiv
\int 
dz |\phi(x, y = 0, z)|^2$
of the binary  $^{168}$Er-$^{164}$Dy BEC, for (a) $N$(Er) = 10000, $N$(Dy) = 20000, and 
$a_{12}=110a_0 $; (b)  $N$(Er) = 10000, $N$(Dy) = 20000, and 
$a_{12}=125a_0$;  (c)  $N$(Er) = 1000, $N$(Dy) = 20000, and 
$a_{12}=130a_0$;  (d)  $N$(Er) = 10000, $N$(Dy) = 500, and 
$a_{12}=160a_0.$ For the trap parameters of this study the length scale 
$l_0={0.5}$ $\mu$m.
} \label{fig3}
\end{center} \end{figure} %

\begin{figure}[!t] 
\begin{center} \includegraphics[width=.49\linewidth]{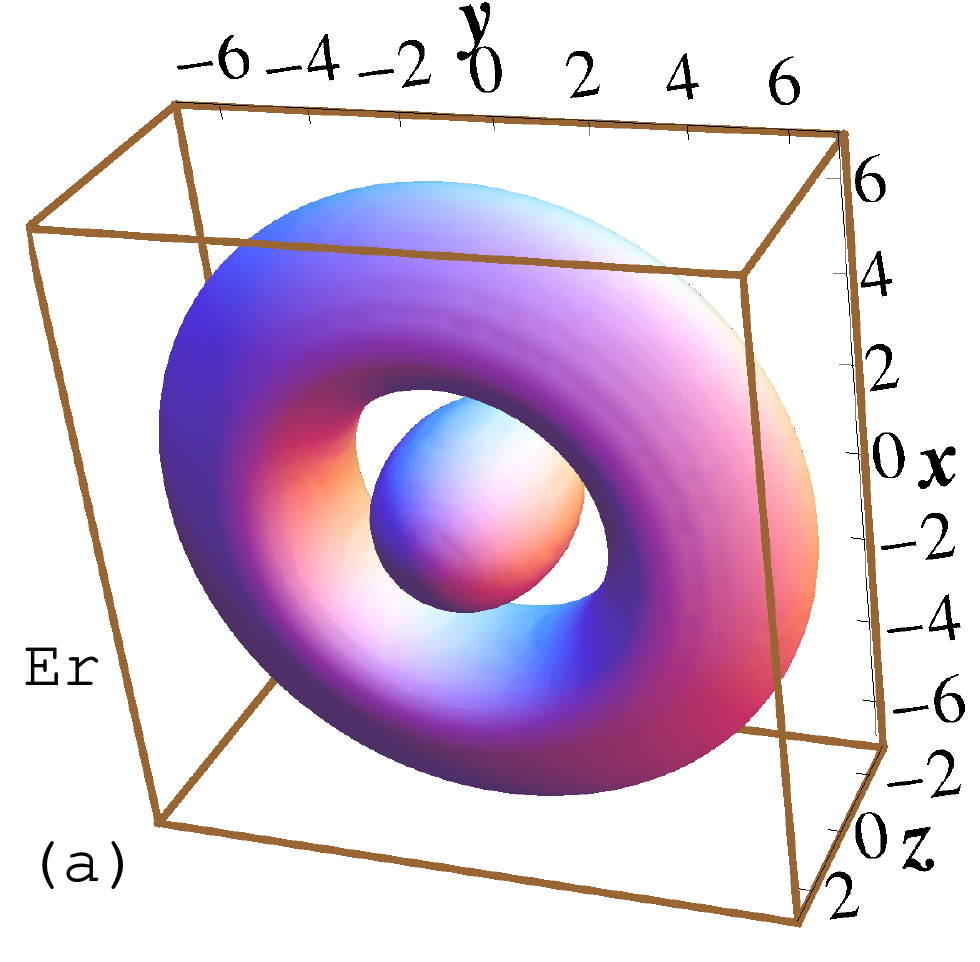} 
\includegraphics[width=.49\linewidth]{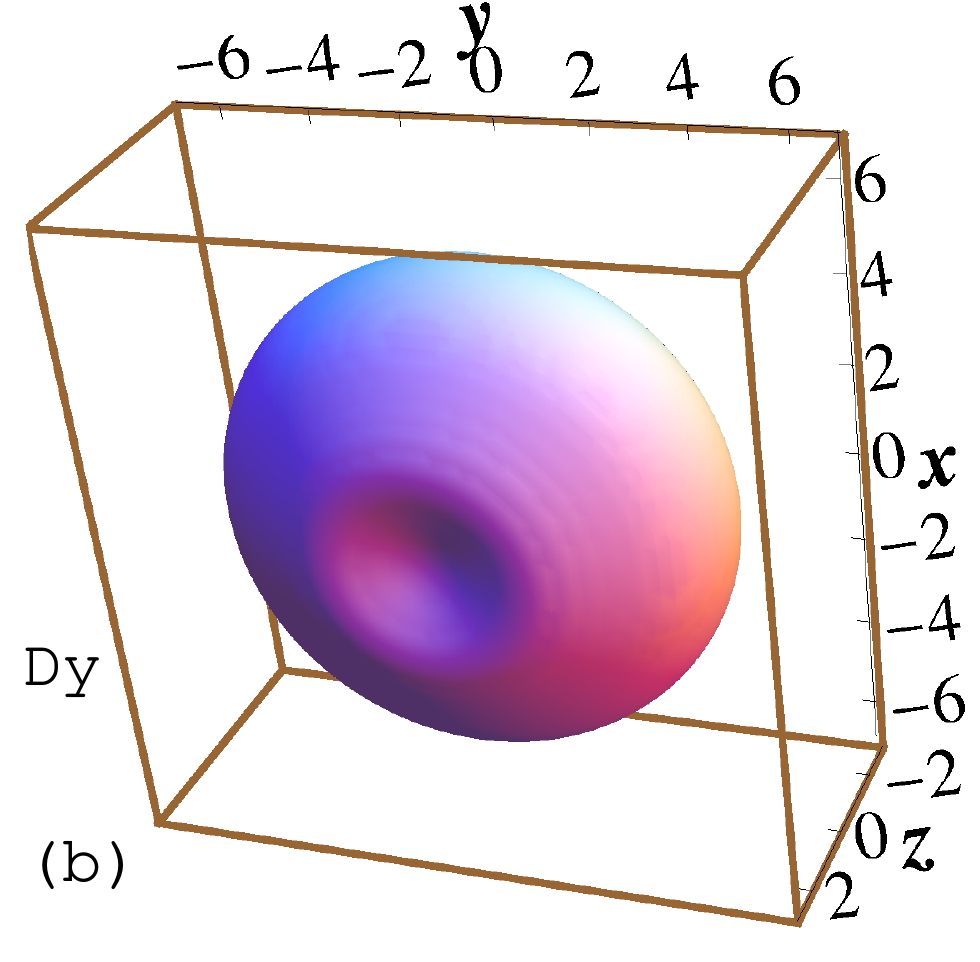} 
\includegraphics[width=.49\linewidth]{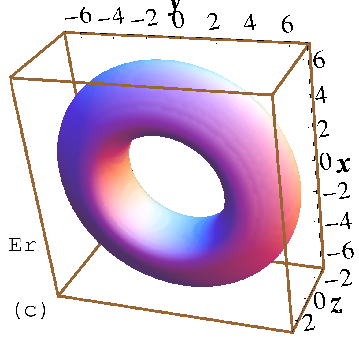} 
\includegraphics[width=.49\linewidth]{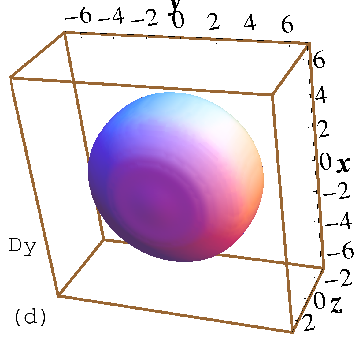} 
\caption{(Color online) Three-dimensional contour plot of the (a) $^{168}$Er and (b) 
$^{164}$Dy densities $|\phi_i(x,y,z)|^2$
of the $^{168}$Er-$^{164}$Dy binary BEC of Fig. 
\ref{fig3} 
(b)  for $a_{12} = 125a_0$. 
The same for the (c) $^{168}$Er and (d) $^{164}$Dy densities of 
the binary BEC of Fig. 
\ref{fig3} (c) for $a_{12} = 
130a_0$.  Density on contour = 0.0005, and 
$N$(Er) = 10000, $N$(Dy) = 20000. The lengths $x,y,z$ are in units 
of $l_0\equiv {0.5}$ $\mu$m. See text for actual values of density inside  and on the surface of the plots. 
}\label{fig4}
\end{center}
\end{figure}

\subsection{Mixing, demixing, and structure formation}

Next we study in detail the density of the two species of atoms in the 
binary dipolar BEC to look for mixing (overlapping phases of two 
components), demixing (separated phases) and distinct  structure formation in the 
shaded regions of the phase plots shown in Fig. \ref{fig2}. For small $a_{12}$ 
the inter-atomic contact repulsion between the two species is small and 
one has a mixed configuration.  As $a_{12}$ is increased the 
inter-species contact repulsion increases and for a large enough value 
of $a_{12}$ one can have demixing of the two species. In the demixed 
configuration one of the species occupy the central region and the other 
the external region. To illustrate this phenomenon, we plot in Fig. 
\ref{fig3} 
the reduced 2D density in the radial direction $|\Phi(x,y=0)|^2= \int dz 
|\phi(x,y=0,z)|^2$ versus $x$ under different situations in case of the 
binary $^{168}$Er-$^{164}$Dy mixture.  The mixing-demixing is 
illustrated in Figs. \ref{fig3} (a), (b), and (c) for 10000 $^{168}$Er and 20000 
$^{164}$Dy atoms, respectively, for $a_{12}=110a_0, 125a_0$, and 
$130a_0$. In Fig. 
\ref{fig3} (a) we have a mixed (overlapping) configuration and 
with the increase of inter-species contact repulsion, in Fig. \ref{fig3} (c) we 
have a demixed (separated) configuration of the two species. In Fig. \ref{fig3}
(b), for $a_{12}=125a_0$ we have a partially demixed configuration. In 
the demixed configuration in Fig. 
\ref{fig3} (c), 20000 $^{164}$Dy atoms have a 
larger contribution to energy of the system and the $^{164}$Dy atoms 
expel the $^{168}$Er atoms from the central region due to inter-species 
repulsion. The opposite panorama is also possible for a larger number of 
$^{168}$Er atoms, while the $^{168}$Er atoms can expel the $^{164}$Dy 
atoms from the central region. Such a situation is illustrated in Fig.  \ref{fig3}
(d) with 10000 $^{168}$Er atoms and 500 $^{164}$Dy atoms for 
$a_{12}=160a_0$. In this case $^{168}$Er atoms have a larger 
contribution to the energy of the system and expel the $^{164}$Dy atoms 
from the center.

A careful examination of the passage from the partially mixed configuration of Fig. \ref{fig3} (b) to the demixed configuration of Fig. \ref{fig3} (c) reveal interesting feature. To study this, we consider the 3D profile of the density $|\phi_i(x,y,z)|^2$
of the two species for the cases depicted in Figs. \ref{fig3} (b) and (c). 
In Figs. \ref{fig4} (a) and (b) we plot the 3D contours  of $^{168}$Er  and  
$^{164}$Dy BEC profiles for the parameters of Fig. \ref{fig3} (b). 
The same for the parameters of Fig. \ref{fig3} (c) are illustrated in Figs. \ref{fig4} (c) and (d). 
The density $|\phi(x,y,z)|^2$
on the surface of these plots is $0.0005l_0^{-3}$. {
With $l_0=0.5$ $\mu$m, and $N$(Er) = 10000, 
and $N$(Dy) = 20000 this  leads to atom densities on contour
of 40 $\mu$m$^{-3}$ for $^{168}$Er and of 80 $\mu$m$^{-3}$ for $^{164}$Dy. The maxima of atom density 
in the interior of the BECs of Figs. \ref{fig4} (a) $-$ (d) are, respectively,
800 $\mu$m$^{-3}$, 1160 $\mu$m$^{-3}$, 460 $\mu$m$^{-3}$, and 1000 $\mu$m$^{-3}$.}
In Figs. \ref{fig4}
 (a) and (b)
the  $^{168}$Er and 
$^{164}$Dy BECs have Saturn-ring-like  and red-blood-cell-like biconcave profiles, respectively.  The 
biconcave profile is a manifestation of the dipolar interaction: the net dipolar repulsion in  $^{164}$Dy in disk shape drives the dysprosium atoms to peripheral region thus creating the biconcave shape as in the single-component 
dipolar BEC. In the single-component case the biconcave profile near the 
stability  limit has been related to roton instability \cite{bohn}.
The Saturn-ring-like $^{168}$Er profile in Fig. \ref{fig4} (a)  is a manifestation of the inter-species dipolar  interaction: the biconcave 
density distribution of the  $^{164}$Dy species with a density minimum at center, due to inter-species repulsion, expels a part of the  $^{168}$Er BEC to the central region of lower density
and another part to the peripheral region, also of lower density, thus creating the Saturn-ring-like profile. Such a profile is not possible in a single-component dipolar BEC, or in a binary BEC without dipolar interaction. 
With further increase of inter-species contact repulsion, for $a_{12}=130a_0$, the contact repulsions dominate over the dipolar interactions and hence plays a major role thus creating a simpler situation of demixing shown in Fig. \ref{fig3} (c) and Figs. \ref{fig4} (c) and (d), where the net inter-species repulsion is so strong that all $^{168}$Er atoms are driven to the peripheral region in the form of a ring with the dominant $^{164}$Dy atoms occupying the central region. 
The ring-like structure in Fig. \ref{fig4} (a) 
as opposed to a shell-like demixed configuration is a consequence of the dipolar interaction: a pure inter-species contact repulsion would have led to a hollow shell-like configuration for   $^{168}$Er  surrounding a compact disk-shaped  $^{164}$Dy core. The dipolar interaction is attractive in the $z$ direction and this transforms the shell-like configuration into a ring. Such transition from a shell-like configuration to a ring with the increase of dipolar interaction has been demonstrated in a single-component dipolar BEC
in a shell-like trap
\cite{adhiring}.

From Figs. \ref{fig3} we find that for a binary dipolar BEC composed of 10000 
$^{168}$Er atoms and 20000 $^{164}$Dy atoms the transition from a mixed 
to a demixed configuration happens around $a_{12}=120a_0$. We verified 
that similar transition also takes place in a binary dipolar BEC for 
10000 $^{168}$Er atoms and a reduced number of $^{164}$Dy atoms also for 
$a_{12} \approx 120a_0$. Nevertheless, the biconcave structure in the 
density of $^{164}$Dy is obtained for a small shaded region below the 
stability line in Fig. \ref{fig2} (a)  with large number of $^{164}$Dy atoms. As 
in the single component case \cite{bohn}, the binary dipolar BEC 
possessing biconcave shape in density in the $^{164}$Dy BEC in the shaded region in 
Fig. \ref{fig2} (a)  is usually  transient  and instability appears 
after crossing the stability lines in Fig. \ref{fig2}. In the case 
of a single-component dipolar BEC
 a conveniently chosen initial density could be necessary for obtaining 
the biconcave structure in density. In imaginary-time propagation the 
initial density should either be chosen as a Gaussian with widths 
slightly larger than the actual widths of the desired state. 
The present structures were obtained by using imaginary 
time propagation of the coupled mean-field equations with Gaussian 
profiles for initial densities. { We started with the Gaussian profiles
of the linear oscillator states and increased the dipolar and nondipolar nonlinearities 
in small steps during numerical simulation until the final nonlinearities are reached. 
In this fashion the structure in density  in the shaded regions of phase plots in Fig. 
\ref{fig2} 
are   obtained.}  

\begin{figure}[!t] 
\begin{center} \includegraphics[clip,width=\linewidth]{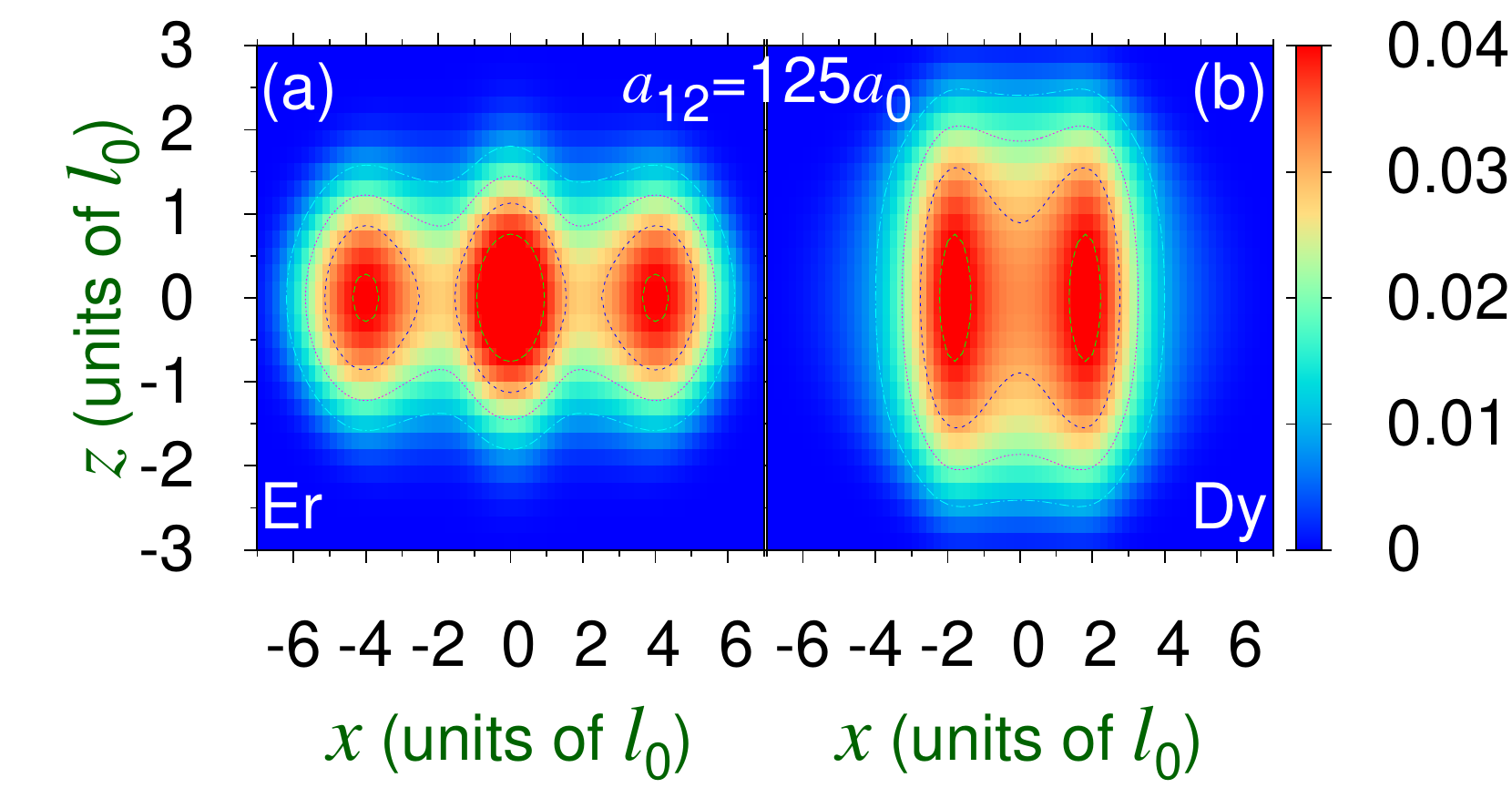} 
\includegraphics[clip,width=\linewidth]{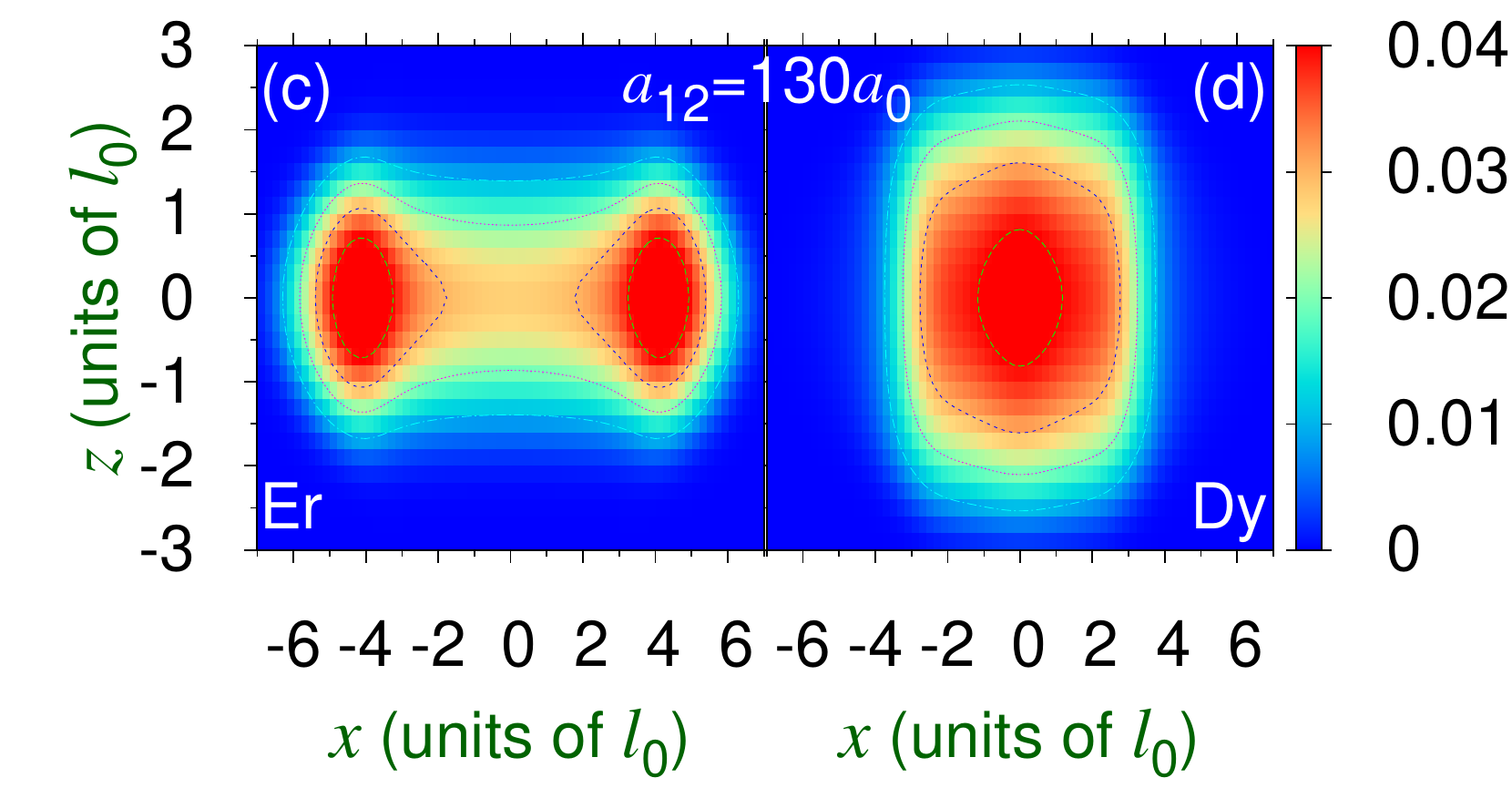} 
\caption{(Color online) Contour plot of 2D density $|\Phi(x,z)|^2\equiv \int 
dy |\phi(x,y,z)|^2$ of the binary $^{168}$Er-$^{164}$Dy BEC for $N$(Er) = 10000, 
$N$(Dy) = 20000 with  $a_{12}=125a_0$ for (a) $^{168}$Er,  (b) $^{164}$Dy, and with  $a_{12}=130a_0$ for 
(c) $^{168}$Er,  and 
(d) $^{164}$Dy. The lengths are expressed in units of $l_0 
(={0.5}$ $\mu$m) 
and the 2D density in units of $l_0^{-2}.$
}\label{fig5}
\end{center}
\end{figure}

\begin{figure}[!t] 
\begin{center} \includegraphics[width=\linewidth]{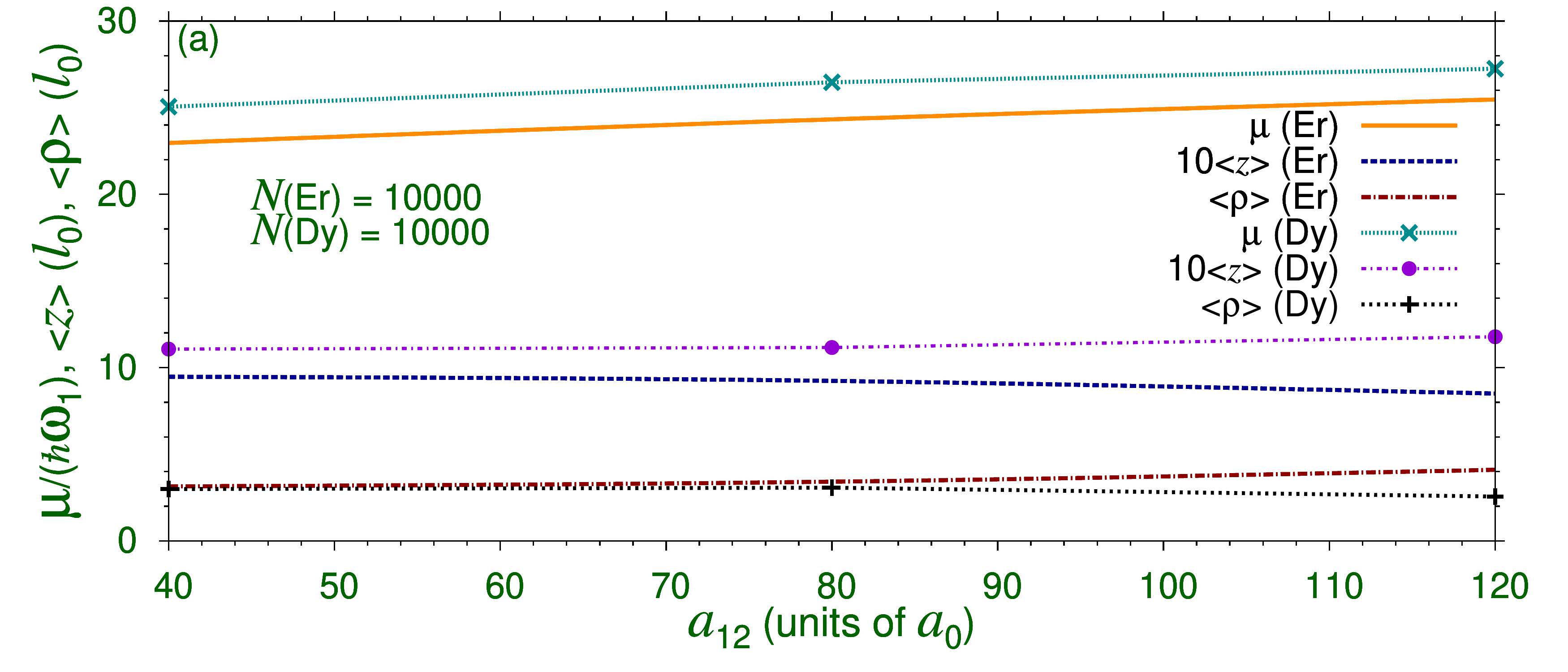} 
\includegraphics[width=\linewidth]{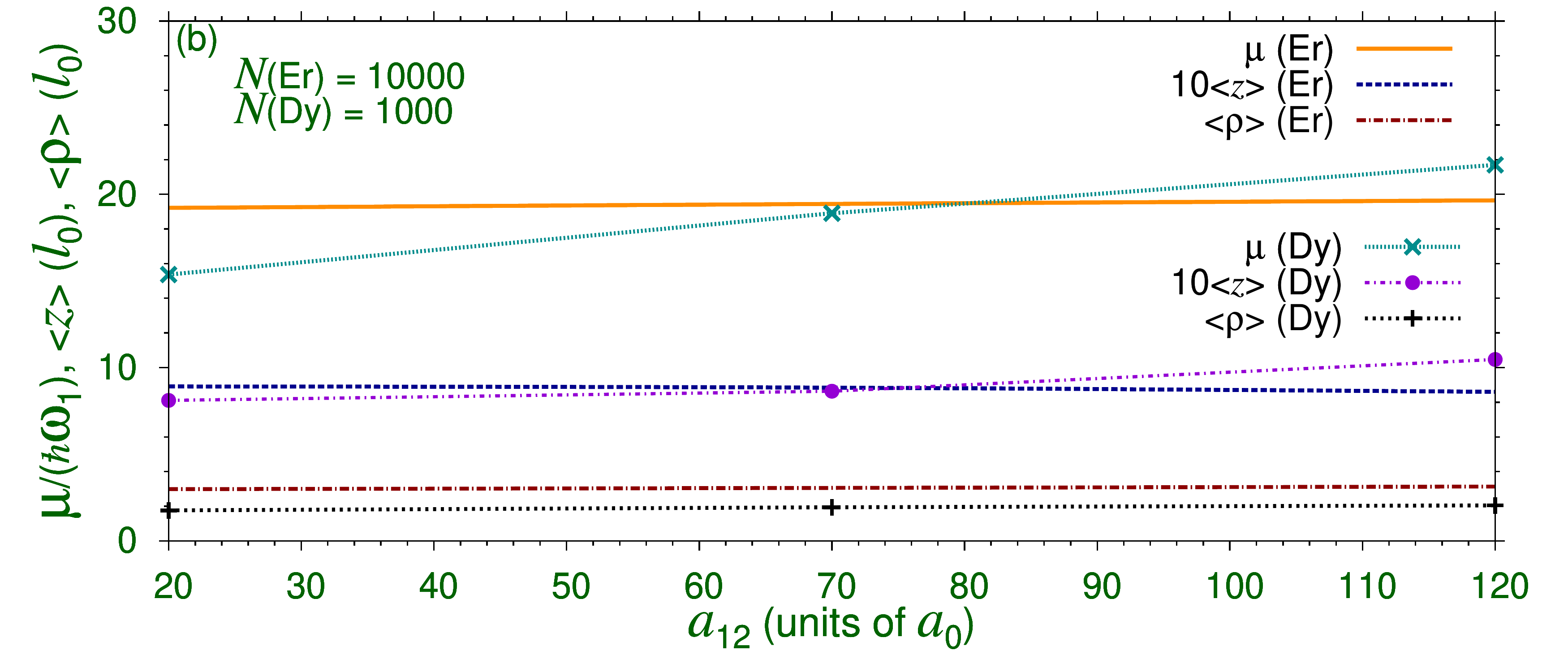} 
\includegraphics[width=\linewidth]{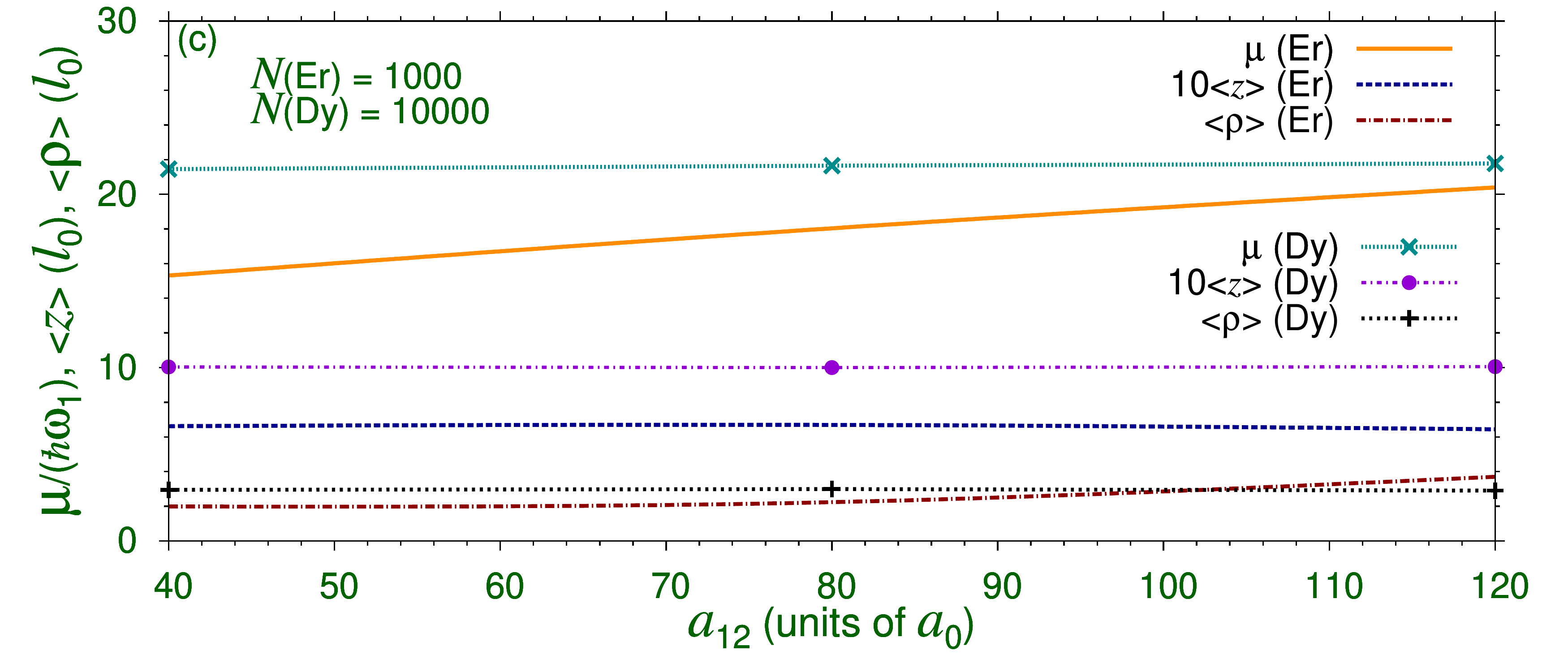} 
\caption{(Color online) Chemical potential $\mu$  and rms sizes $\langle z \rangle $, 
$\langle \rho \rangle $ versus inter-species scattering length for a
binary $^{168}$Er-$^{164}$Dy BEC for (a) $N$(Er) = $N$(Dy)
 = 10000; 
(b) $N$(Er)  
=10000,  $N$(Dy)
=1000;
 (c) $N$(Er) 
= 1000,  $N$(Dy)
= 10000. The unit of length is $l_0= {0.5} $ $\mu$m.}\label{fig6}
\end{center}
\end{figure}

To gain further insight into the density distributions in Figs. \ref{fig4} (a) $-$ (d) 
we plot in Figs. \ref{fig5} (a) $-$ (d) contour plots of asymmetric 2D density $|\Phi(x,z)|^2
=\int dy |\phi(x,y,z)|^2$ of the binary $^{168}$Er-$^{164}$Dy mixture for parameters corresponding to Figs. \ref{fig4} (a) $-$ (d), respectively.  The double and single saddle structures in density in the $x-z$ plane in Figs. \ref{fig5} (a) and (b) are  manifestations of the dipolar interaction. In these plots a minimum in density along $x$ direction coincides with a maximum in density in $z$ direction thus creating a saddle point, which appears as a clear manifestation of the 
saddle-shaped dipolar interaction \cite{saddle}. Such a density distribution is not possible in the absence of dipolar interaction and to the best of our knowledge has not been demonstrated even in the single-component dipolar BEC. 
 In Fig. \ref{fig5} (a), for $a_{12}=125a_0$, we have two saddle points on the 
$x$ axis in the density of $^{168}$Er and one saddle point in the 
density of $^{164}$Dy. Similarly, for $a_{12}=130a_0$, we have a single 
saddle point only in the density of $^{168}$Er and none in $^{164}$Dy. 
These saddle points have curvatures consistent with saddle-shaped 
dipolar interaction \cite{saddle} with negative curvature along the 
direction of magnetization ($z$ axis) and positive curvature orthogonal 
to it ($x$ axis).

In Fig. 6 we plot the chemical potential $\mu$ and root mean square (rms) sizes 
$\langle z \rangle$, and $\langle \rho \rangle$ versus inter-species 
scattering length $a_{12}$, for 
(a) $N$(Er) = $N$(Dy)
 = 10000; 
(b) $N$(Er)
 = 10000,  $N$(Dy)
 =1000;
 (c) $N$(Er) 
 = 1000,  $N$(Dy)
 = 10000. It is found that the chemical potential 
and rms sizes are slowly varying functions of inter-species scattering length. 
As expected, the chemical potential of the first species increases more rapidly with 
$a_{12}$ when there are more atoms in the second species and vice versa.

\begin{figure}[!t] 
\begin{center} \includegraphics[width=\linewidth,clip]{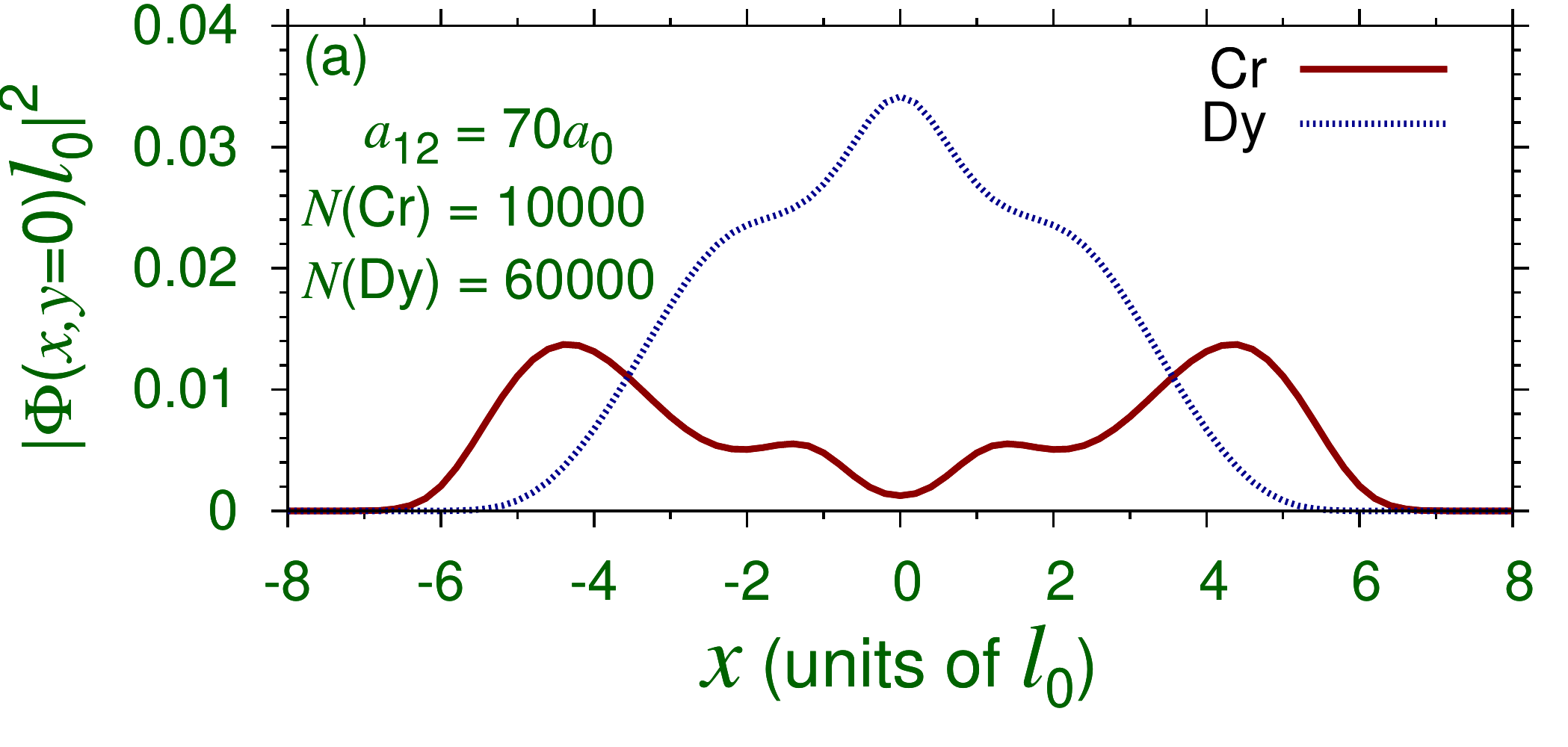} 
\includegraphics[width=\linewidth,clip]{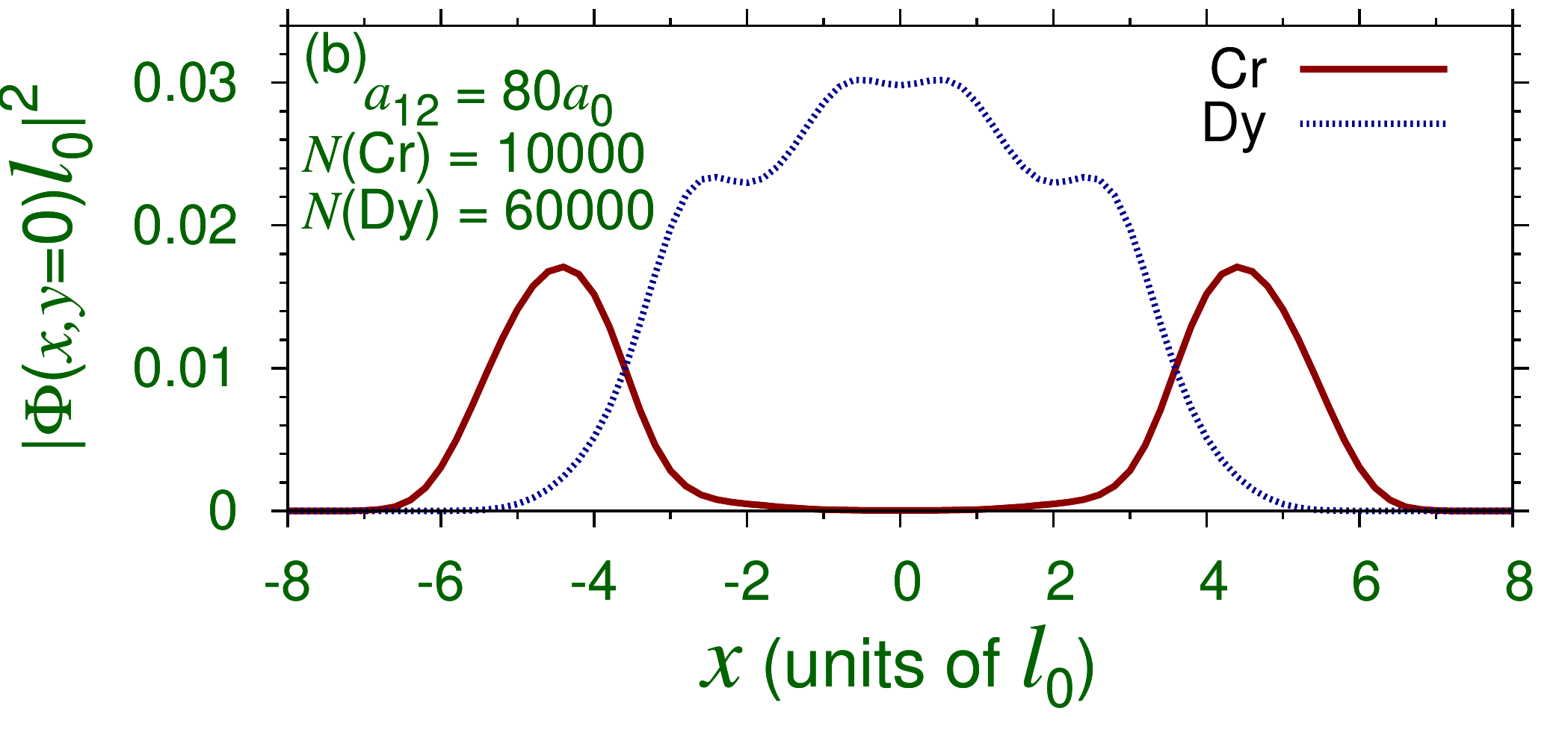} 
\includegraphics[width=.49\linewidth,clip]{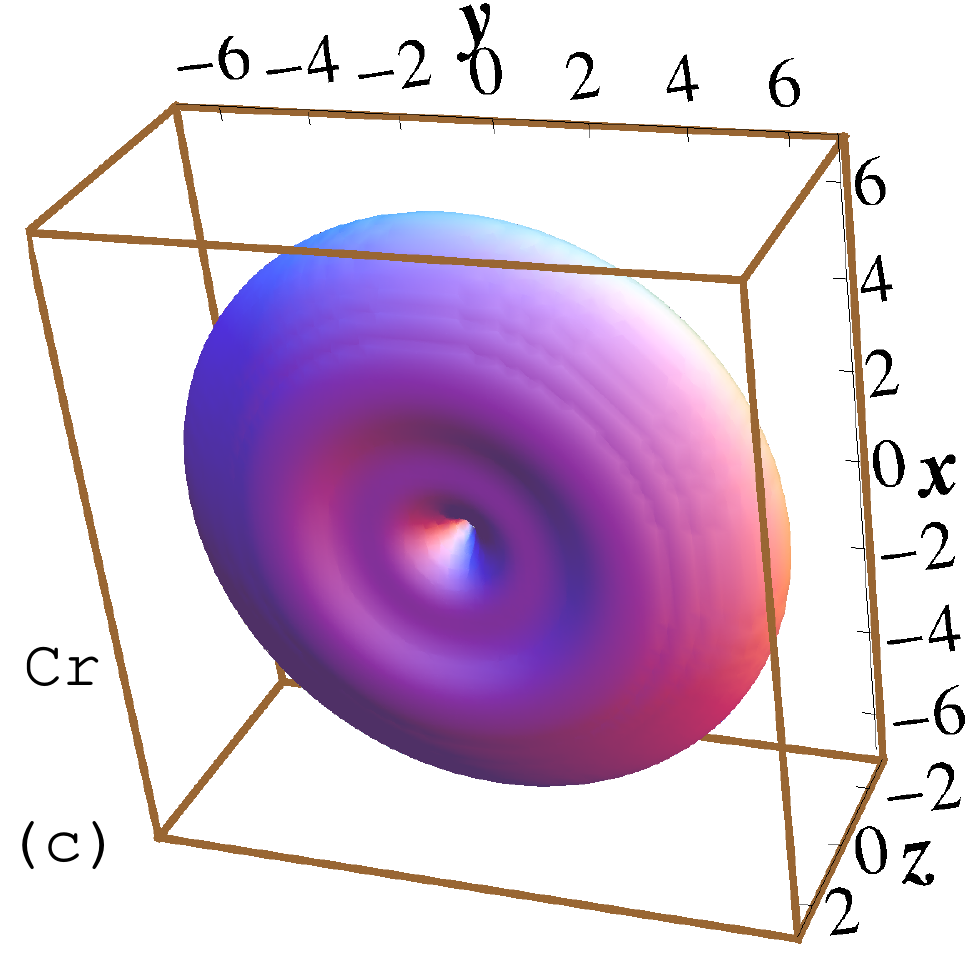} 
\includegraphics[width=.49\linewidth,clip]{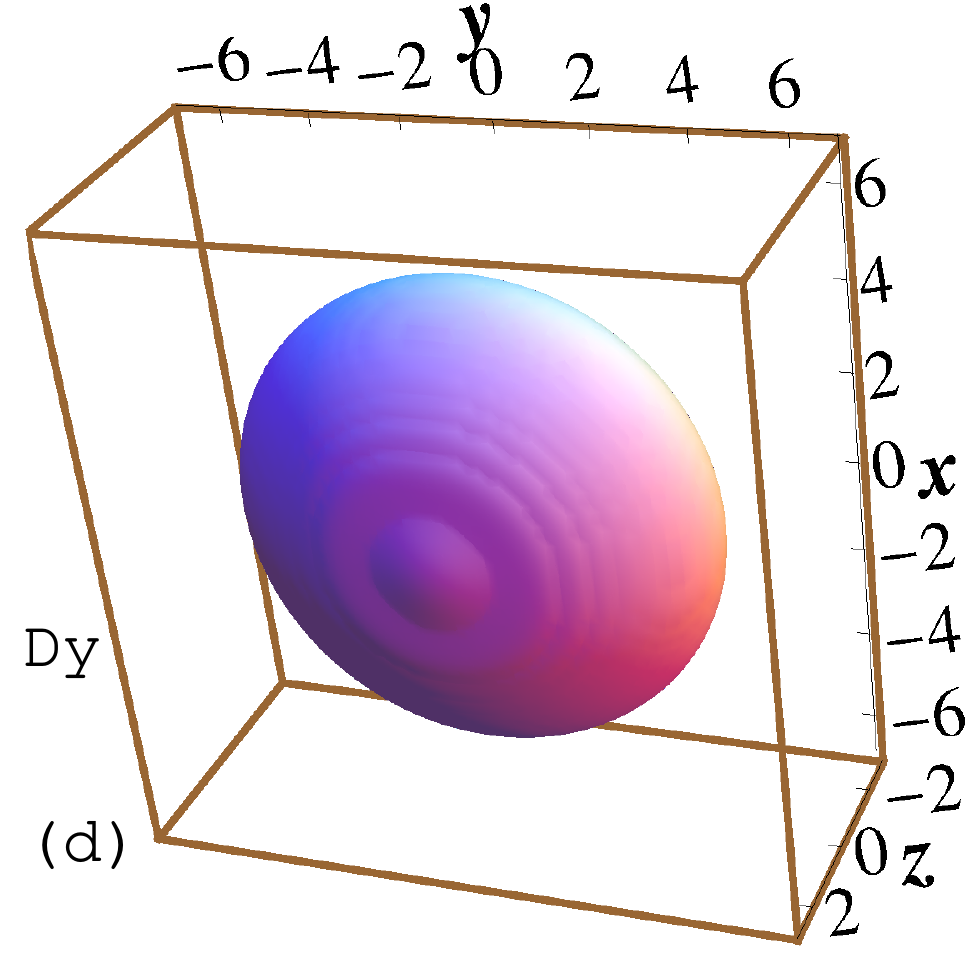}
\includegraphics[width=.49\linewidth,clip]{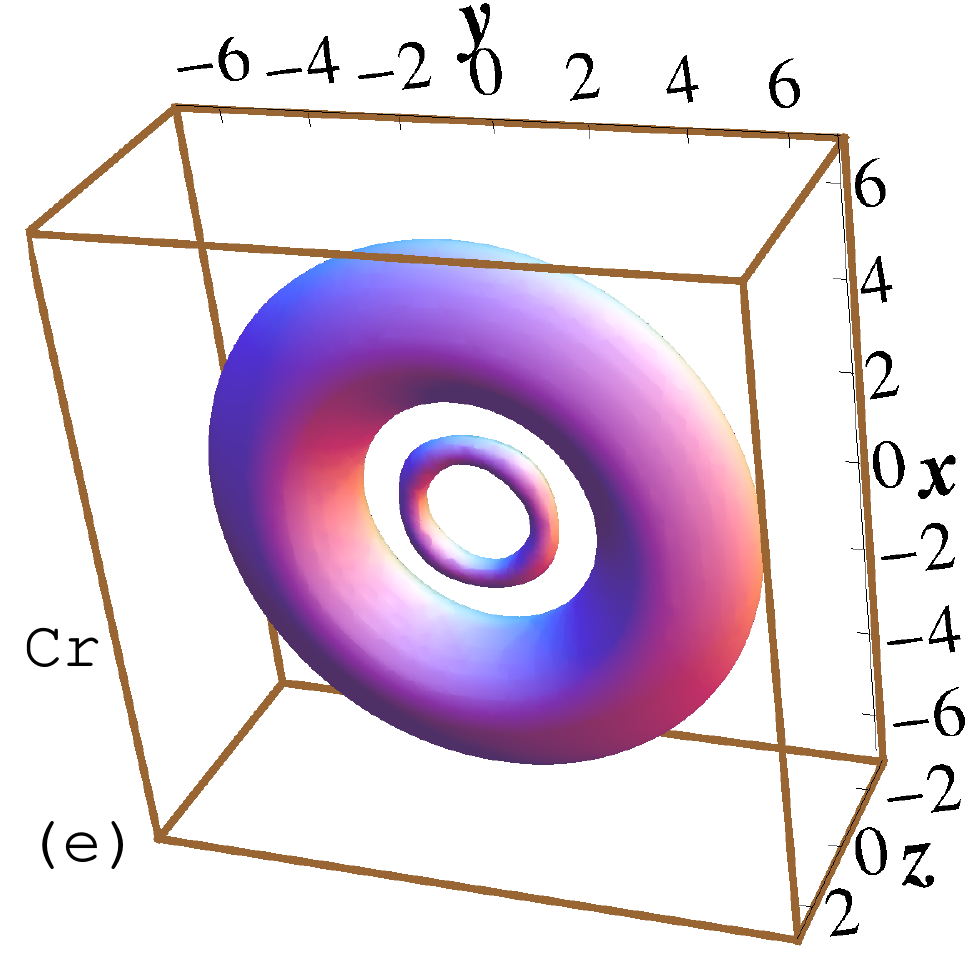} 
\includegraphics[width=.49\linewidth,clip]{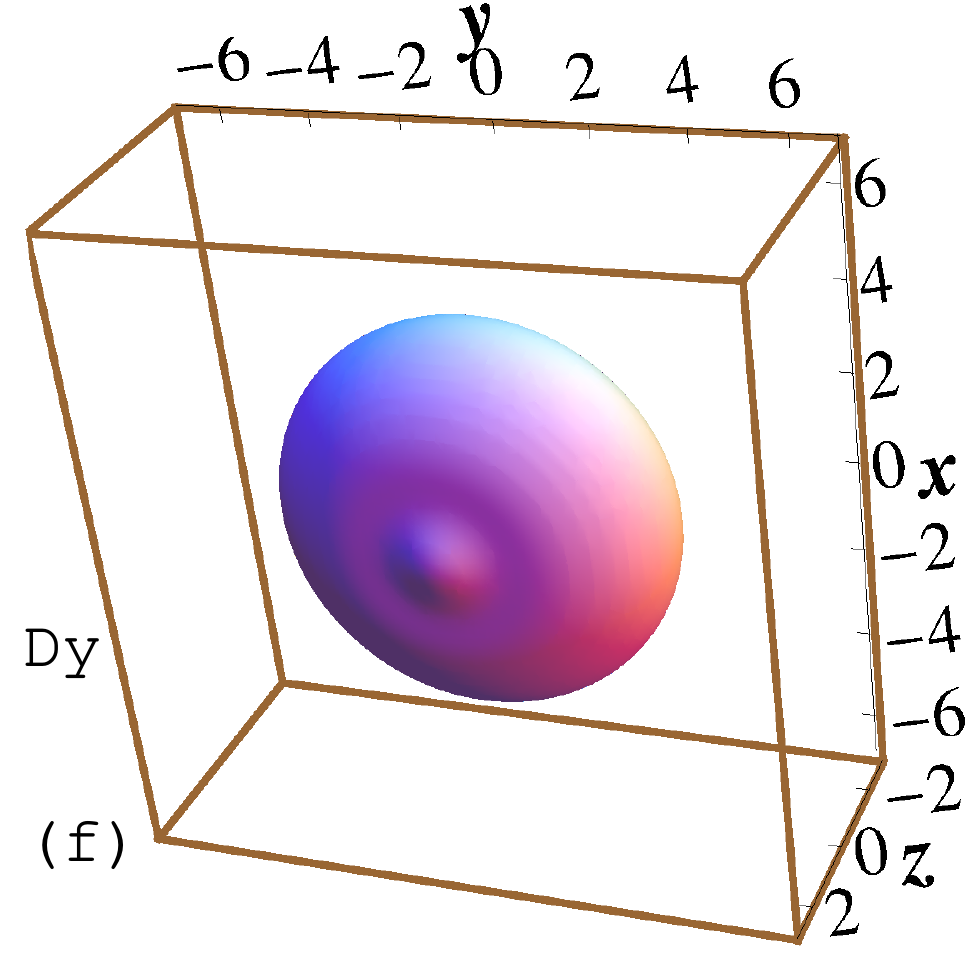}
\caption{(Color online) Two-dimensional radial density along  $ x$ 
axis $|\Phi(x, y = 0)|^2 \equiv
\int 
dz |\phi(x, y = 0, z)|^2$
of the binary  $^{52}$Cr-$^{164}$Dy BEC, for   (a) $a_{12}=70a_0 $, (b)   
$a_{12}=80a_0$. Three-dimensional contour plot of the (c) $^{52}$Cr and (d) 
$^{164}$Dy densities $|\phi_i(x,y,z)|^2$ of the binary BEC of Fig. 3 
(a)  for $a_{12} = 70a_0$  with the cut-off density on contour of 0.0005.  
The same densities of (e) $^{52}$Cr and (f) 
$^{164}$Dy with the cut-off density on contour of 0.0025.  
In all cases
$N$(Cr) = 10000, $N$(Dy) = 60000.
The lengths $x,y,z$ are in units      
of $l_0 = {1}$ $\mu$m.
}\label{fig7}
\end{center}
\end{figure}

With this study of the binary $^{168}$Er-$^{164}$Dy BEC, we also 
considered a detailed investigation of the binary $^{52}$Cr-$^{164}$Dy 
BEC. Similar mixing, demixing, and structure formation are also found in 
this second case, although the quantitative results and estimates are 
different. We only report here the studies on the interesting structure 
formation in the binary $^{52}$Cr-$^{164}$Dy BEC. Indeed we find that 
one can have the condensate with biconcave shape in this case.  In Figs. 
\ref{fig7} (a) and (b) we plot the linear density in the $x-y$ plane along the 
$x$ direction $|\Phi(x,y=0)|^2$, as in Fig. \ref{fig3}, for a binary mixture 
of 10000 $^{52}$Cr atoms and 60000 $^{164}$Dy atoms for 
$a_{12}=70a_0$ and $80a_0$, respectively. Figure \ref{fig7} (a) illustrates a 
case similar to Fig. \ref{fig3} (b) denoting a transition from a mixed state to a 
demixed state. For $a_{12}< 60a_0$, we have a fully mixed configuration 
of the two condensates as in Fig. \ref{fig3} (a). For $a_{12}=70a_0$ a biconcave 
shape has appeared in the density of $^{52}$Cr.  For $a_{12}> 75a_0$, a 
complete demixed configuration appears as illustrated in Fig. \ref{fig7} (b) for 
$a=80a_0$, where the $^{52}$Cr atoms are expelled from the central 
region occupied by only $^{164}$Dy atoms.  In Figs. \ref{fig7} (c) and (d) we 
show the 3D contour of the density $|\phi(x,y,z)|^2$ of
$^{52}$Cr and $^{164}$Dy BECs corresponding 
to the parameters of Fig. \ref{fig7} (a) with the  density 0.0005$l_0^{-3}$ 
on contour. {This corresponds to number densities on the contour 
of  5 $\mu$m$^{-3}$ and 30 $\mu$m$^{-3}$ in $^{52}$Cr
and $^{164}$Dy, respectively, for $l_0=1$ $\mu$m.} 
The $^{52}$Cr BEC shows a pronounced 
biconcave shape which is a direct manifestation of the dipolar interaction.
However, we could not find a biconcave shape in the density of the 
$^{164}$Dy BEC, as in the case of a binary dipolar $^{168}$Er-$^{164}$Dy
mixture,
possibly because the inter-species dipole interaction in  
the present system is much smaller in strength compared to the 
same in the binary $^{168}$Er-$^{164}$Dy system so as to create the 
biconcave shape in density in the presence of dominating large contact repulsions acting on $^{164}$Dy. 
The much larger dipole interactions in $^{164}$Dy might be 
responsible for the biconcave  shape in density in Fig. 4 (b). 
Even in the single-component case, the biconcave profile in density 
appears for  certain values of the trap aspect ratio \cite{bohn}, and it is also possible that for other values of trap aspect ratio the biconcave profile might appear in the density of  $^{164}$Dy in the present binary $^{52}$Cr-$^{164}$Dy mixture. However, the dipolar interaction leaves its signature in a different fashion 
as can be seen in Figs. \ref{fig7} (e) and (f), where we plot the same densities of Figs. \ref{fig7} 
(c) and (d) but now with the density 0.0025$l_0^{-3}$ on the contour. 
 {This corresponds to number densities on the contour 
of  25 $\mu$m$^{-3}$ and 150 $\mu$m$^{-3}$ in $^{52}$Cr
and $^{164}$Dy, respectively, for $l_0=1$ $\mu$m.}
The Figs. \ref{fig7} 
(e) and (f) thus show the structure in the interior of the BECs illustrated in Figs. \ref{fig7} 
(c) and (d). Along the radial direction in the $x$-$y$ plane,  we find in 
Figs. \ref{fig7} (e) and (f) that  the maximum of density 
of one component is accompanied by the minimum of density of the other component due
to interspecies contact repulsion and the density of one component may have several 
local maxima. {The global maxima in densities of Figs. \ref{fig7} (e) and (f)
are 0.0064$l_0^{-3}$ and 0.013$l_0^{-3}$, respectively, corresponding to number densities of 64 $\mu$m$^{-3}$ and 780
$\mu$m$^{-3}$ for  $^{52}$Cr
and $^{164}$Dy.}

\begin{figure}[!t] 
\begin{center} \includegraphics[width=\linewidth,clip]{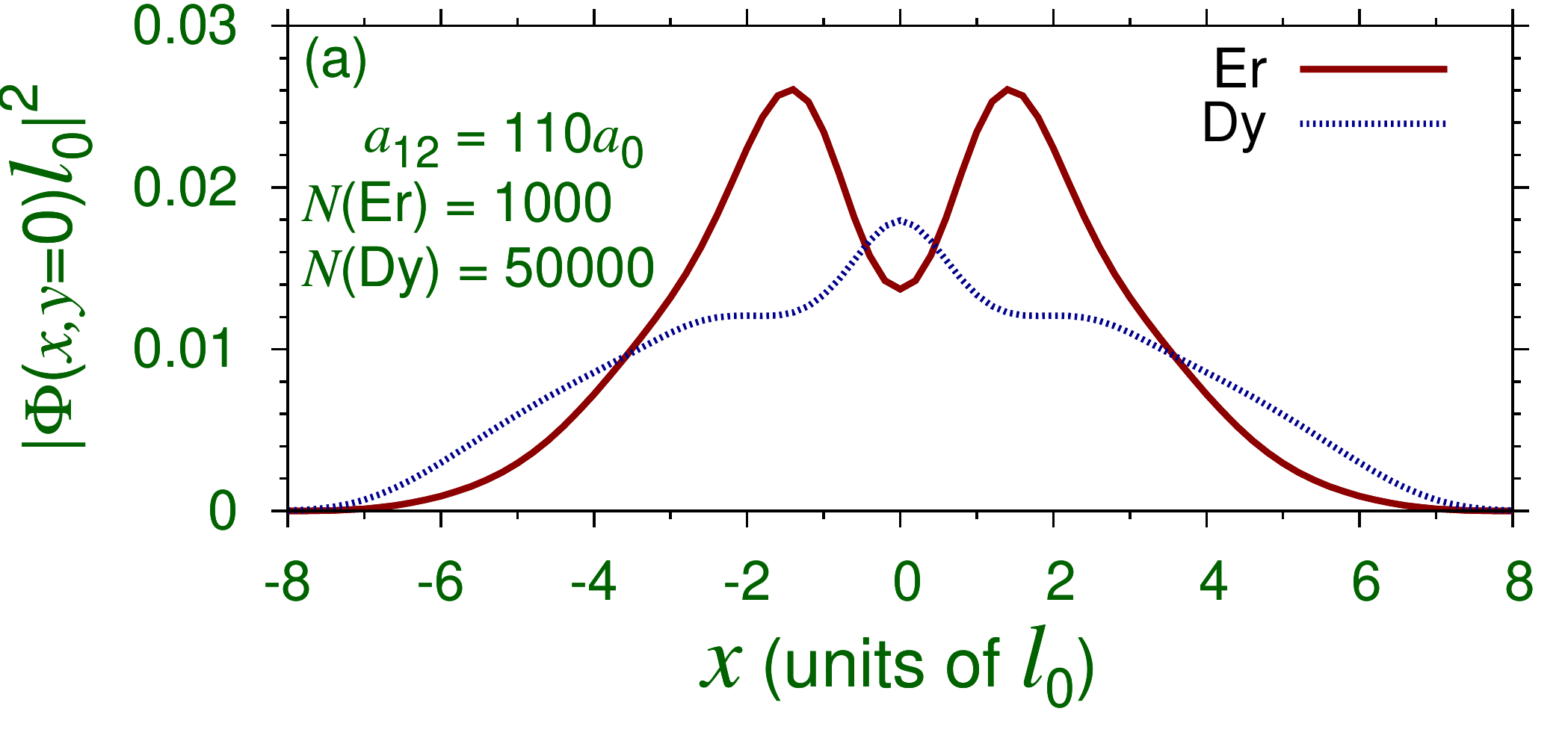} 
\includegraphics[width=.49\linewidth,clip]{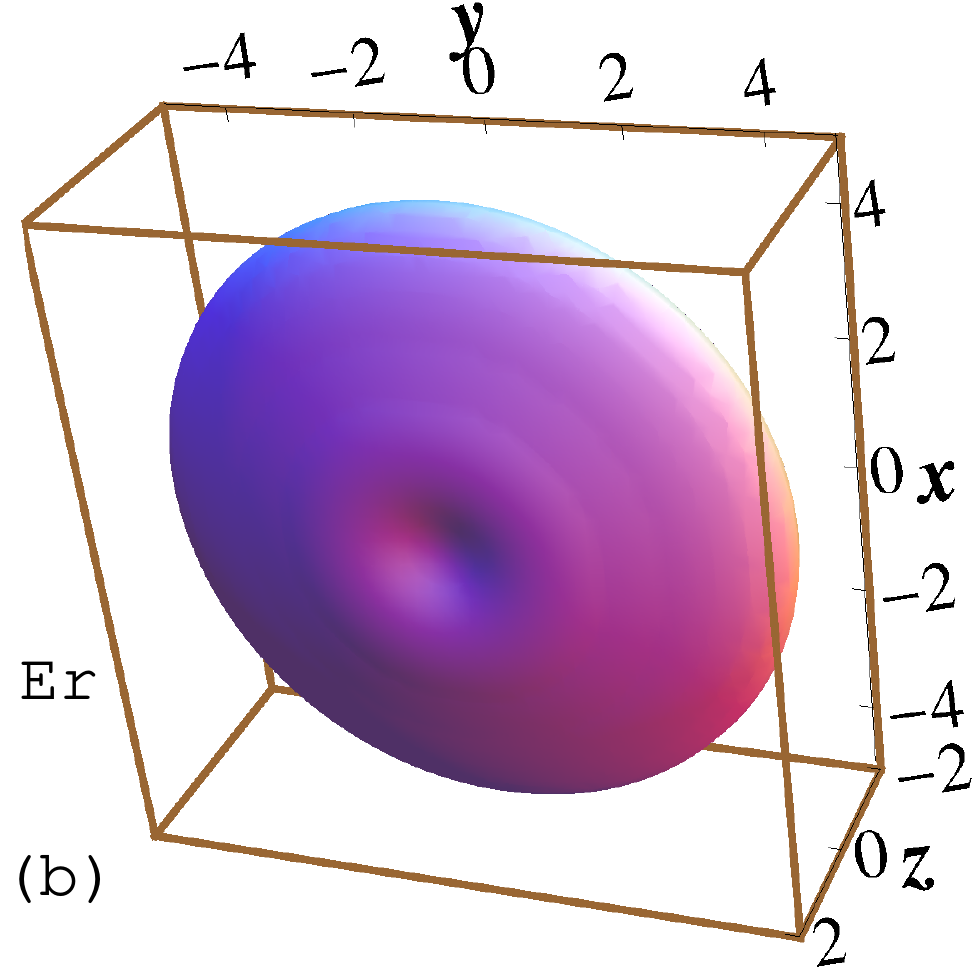} 
\includegraphics[width=.49\linewidth,clip]{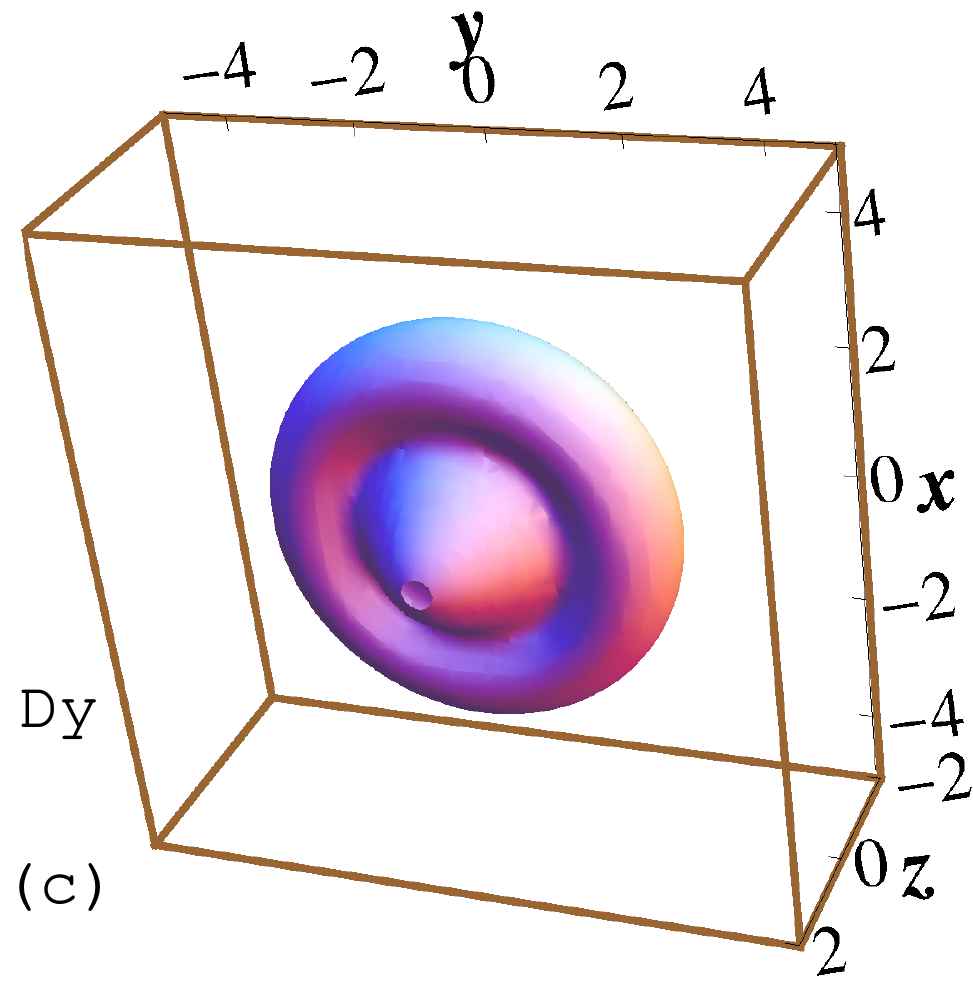}

\caption{(Color online) Two-dimensional radial density along  $ x$ 
axis $|\Phi(x, y = 0)|^2 \equiv
\int 
dz |\phi(x, y = 0, z)|^2$
of the binary  $^{168}$Er-$^{164}$Dy BEC of $N$(Er) = 1000, $N$(Dy) = 50000,
 for   (a) $a_{12}=110a_0 $. 
Three-dimensional contour plot of the (b) $^{168}$Er and (c) 
$^{164}$Dy density $|\phi_i(x,y,z)|^2$ of the binary BEC of  
(a). 
  Density on 3D contour = 0.0025.
The lengths $x,y,z$ are in units      
of $l_0 = {0.5}$ $\mu$m.
}\label{fig8}
\end{center}
\end{figure}

{ Finally, we study the distinct structure in densities of
binary dipolar BECs with a smaller number $^{168}$Er atoms  
in the shaded region of 
Fig. \ref{fig2} (b). In particular we consider the partially mixed configuration with 
$a_{12}=110a_0, N$(Er)
= 1000, $N$(Dy)
= 50000.  In Fig. \ref{fig8} (a), we illustrate the 2D  radial density along  $ x$ 
axis $|\Phi(x, y = 0)|^2 \equiv
\int 
dz |\phi(x, y = 0, z)|^2$
of this binary  $^{168}$Er-$^{164}$Dy BEC. In Figs. \ref{fig8} (b) and (c) the contour plots of the 
corresponding 3D 
densities  of $^{168}$Er and  $^{164}$Dy are shown
{with a cut-off density 0.0025$l_0^{-3}$ on the contour.
The maximum densities inside the $^{168}$Er and  $^{164}$Dy BECs are 0.014$l_0^{-3}$ and 0.0038$l_0^{-3}$, respectively.}
 In this case a biconcave shape has appeared 
in the density profile of $^{168}$Er BEC in Fig. \ref{fig8} (b). A Saturn-anel-like profile in the density 
of  $^{164}$Dy BEC is shown in Fig. \ref{fig8} (c). The central high-density region of $^{164}$Dy BEC has 
expelled the $^{168}$Er BEC from the central region, thus creating a biconcave shape in density. 
If we compare Figs. \ref{fig4} (a) and (b) with Figs. \ref{fig8} (b) and (c), we find that the 
roles of  $^{164}$Dy and $^{168}$Er BECs have interchanged. In Fig. \ref{fig4} (a), the $^{168}$Er BEC
has a Saturn-anel-like profile, whereas, in Fig. \ref{fig8} (c), the $^{164}$Dy BEC has similar profile. 
These profiles are a consequence of dipolar interaction. 
}

\section{Summary }
 
Using a mean-field description we studied the static properties of binary
disk-shaped  
dipolar BECs $^{168}$Er-$^{164}$Dy and $^{52}$Cr-$^{164}$Dy  to search for the effect of inter- and intra-species dipolar interactions employing realistic    
values of inter- and intra-species dipolar interactions as of intra-species 
scattering lengths. The yet unknown inter-species  scattering length is considered as a variable parameter.  The binary system is found to be stable for number of atoms below a critical value. The stability domain for the system is illustrated in  convenient phase plots involving number of atoms of the species and the inter-species scattering length $a_{12}$. For 
 small values of $a_{12}$ a mixed state (with overlapping phase) of the two species  emerges and this state transforms into a demixed state (with separated phase) for larger $a_{12}$. Just below the stability line, 
in the region of transition from the mixed to a demixed configuration distinct
structures in 3D densities may appear in the form of the Saturn-ring-like shape  or red-blood-cell-like biconcave shape. Such structures are a direct manifestations of the dipolar interaction.  Similar biconcave shape  in the density of a single-component dipolar BEC was found and related to roton instability, which is presumably also responsible for these structures in the binary dipolar BECs.   
After this investigation was finished we came to know about another recent  work on binary dipolar 
BEC in 2D \cite{binary}.

\acknowledgments
We thank FAPESP  and  CNPq (Brazil)  for partial support.

\end{document}